\documentclass[12pt,a4paper,reqno]{amsart}

\usepackage{figsize,subfigure,ifthen,calc,psfrag}
\usepackage{enumerate,comment}
\usepackage{xspace}
\usepackage{latexsym}
\usepackage{amsfonts}
\usepackage{amsmath}
\usepackage{amssymb}
\usepackage{amsxtra}
\usepackage{hyperref}
\usepackage{amsthm}
\usepackage{graphicx}
\usepackage{multicol}
\usepackage{url}
\usepackage{rotating}
\usepackage[bottom]{footmisc}
\usepackage{harvard}
\usepackage{mathdots}
\usepackage{dsfont}
\usepackage{nameref}
\usepackage{hhline}
\usepackage{verbatim}
\usepackage{lmodern}
\usepackage{alltt}
\usepackage{amsaddr}
\usepackage{mathdots}
\numberwithin{equation}{section}

\allowdisplaybreaks

\begin{document}


\def\tends{\rightarrow}

\def\C{{\mathbb C}}
\def\R {{\mathbb R}}
\def\P {{\mathbb P}}
\def\N{{\mathbb N}}
\def\Z{{\mathbb Z}}
\def\cL{{\mathcal L}}

\newcommand\be{\begin{eqnarray}}
\newcommand\ee{\end{eqnarray}}
\newcommand\bee{\begin{eqnarray*}}
\newcommand\eee{\end{eqnarray*}}
\newcommand\mbf{\mathbf}
\newcommand\un{\underline}

\newcommand{\Proof}{{\sc Proof}\,}

\newtheorem{lemma}{Lemma}[section]
\newtheorem{theorem}[lemma]{Theorem}
\newtheorem{proposition}[lemma]{Proposition}
\newtheorem{corollary}[lemma]{Corollary}

\theoremstyle{definition}
\newtheorem{remark}[lemma]{Remark}
\newtheorem{example}[lemma]{Example}
\newtheorem{definition}[lemma]{Definition}

\theoremstyle{plain}
\newcounter{ASSUMPTION}
\newtheorem{assumption}[ASSUMPTION]{Assumption}

\textheight22.25cm
\textwidth17.5cm

\title[\sf Frequency Domain  Bootstrap]{A Frequency Domain Bootstrap for General Stationary Processes}

\author[\sf M. Meyer]{Marco Meyer}
\address{Technische Universit\"at Braunschweig, Inst. f. Math. Stochastik, Universit\"atsplatz 2, D--38106 Braunschweig, Germany. Email: marco.meyer@tu-bs.de}
\author[\sf E. Paparoditis]{Efstathios Paparoditis}
\address{University of Cyprus,
         Department of Mathematics and Statistics,
         1678 Nicosia,
         Cyprus.}
\author[\sf J.-P. Kreiss]{Jens-Peter Kreiss}
\address{Technische Universit\"at Braunschweig, Inst. f. Math. Stochastik, Universit\"atsplatz 2, D--38106 Braunschweig, Germany. Email: j.kreiss@tu-bs.de}

\date{\today}


\vspace{1cm}

\begin{abstract}  Existing frequency domain methods for bootstrapping time series  have a limited range.
Consider for instance the class of spectral mean statistics (also called integrated periodograms) which includes many important statistics in time series analysis, such as sample autocovariances and autocorrelations among other things.  Essentially, such frequency domain bootstrap procedures cover the case of linear time series with independent innovations, and some even require the time series to be Gaussian.
In this paper we propose a new, frequency domain bootstrap method  which is consistent for a much wider range of stationary processes and can be applied to a large class of periodogram-based statistics. It introduces a new concept of convolved periodograms of smaller samples  which uses    pseudo
periodograms  of subsamples  generated in a way  that correctly   imitates the weak dependence structure of the periodogram. 
We show consistency for this procedure for a general class of stationary  time series, ranging clearly beyond linear processes, and for general spectral means and ratio statistics.  Furthermore, and for the class of spectral means, 
we also show, how,  using this new approach,  existing  bootstrap methods, which  replicate appropriately only the dominant part of the distribution of interest, can be corrected.
The finite sample performance of the new bootstrap  procedure is illustrated via simulations.
\end{abstract}

\maketitle

\setlength{\parindent}{0pt}

\section{Introduction}

Frequency domain bootstrap methods for time series are quite attractive because in many situations they can be successful without imitating the (potentially complicated) temporal dependence structure of the underlying stochastic process, as is the case for time domain bootstrap methods. Frequency domain methods mainly focus on bootstrapping the periodogram which is defined for any time series $ X_1,\ldots, X_n$ by
\be
I_n(\lambda)=\frac{1}{2\pi n} \left| \, \sum_{t=1}^{n} X_t \, e^{-i\lambda t} \, \right|^2, \quad \lambda \in [-\pi,\pi]\,. \label{periodogram}
\ee
The periodogram is an important frequency domain statistic and many statistics of interest in time series analysis can be written as functions of the periodogram. Furthermore, it obeys some nice nonparametric properties for a wide class of stationary processes, which make frequency domain bootstrap methods appealing.
In particular, periodogram ordinates rescaled by the spectral density are asymptotically standard exponential distributed and, moreover, periodogram ordinates corresponding to different frequencies in the interval $(0,\pi)$ are asymptotically independent. This asymptotic independence essentially means that the classical i.i.d.~bootstrap of drawing with replacement, as has been introduced by Efron (1979), can potentially be applied to bootstrap the periodogram, in particular the properly rescaled periodogram ordinates. Motivated by these considerations, many researchers have developed bootstrap methods in the frequency domain which generate pseudo-periodogram ordinates with the intent to mimic the stochastic behavior of the ordinary periodogram.\\
A multiplicative bootstrap approach for the periodogram has been investigated by Hurvich and Zeger (1987), Franke and H\"ardle (1992) and Dahlhaus and Janas (1996). The main idea is to exploit the (asymptotic) independence of the periodogram and to generate new pseudo periodogram ordinates by  multiplying an estimator of the spectral density at the frequencies  of interest with pseudo innovations obtained by an i.i.d.~resampling of appropriately defined frequency domain residuals. Franke and H\"ardle (1992) proved validity of such an approach for estimating the distribution of nonparametric spectral density estimators for linear processes of the form
\be
X_t= \sum_{j=-\infty}^{\infty} a_j \varepsilon_{t-j}, \quad \quad t\in \Z, \label{linearproc}
\ee
where $ (\varepsilon_{t})_{t \in \Z} $ denotes an i.i.d.~white noise process. Shao and Wu (2007) established validity of this procedure for the same statistic but for a much wider class of stochastic processes.\\
However, beyond nonparametric spectral density estimators, the range of validity of this bootstrap approach is limited. In fact, even in the special case of linear processes as given by \eqref{linearproc} with i.i.d.~white noise, Dahlhaus and Janas (1996) showed that the multiplicative approach fails to consistently estimate the limiting distribution of very basic statistics like sample autocovariances. This failure is due to the following: For many periodogram-based statistics of interest consistency is achieved by letting the number of frequencies at which the periodogram is evaluated increase with increasing sample size. The dependence between periodogram ordinates at different frequencies vanishes asymptotically, but the rate of this decay typically just compensates the increasing number of frequencies. This leads to the fact that the dependence structure actually shows up in the limiting distribution of many statistics of interest. Since the bootstrap pseudo-periodogram ordinates generated by the multiplicative approach are independent, this approach fails to imitate the aforementioned dependence structure of periodogram ordinates.\\
As a consequence, beyond the class of nonparametric spectral density estimators, validity of this frequency domain bootstrap approach can be established only for a restricted class of processes and statistics. To be precise, even in the special case of linear processes \eqref{linearproc} with i.i.d.~white noise, the approach works only under additional assumptions, such as Gaussianity of the time series, or for specific statistics such as ratio statistics. For nonlinear processes, even for processes with a linear structure as in \eqref{linearproc} but with non-i.i.d.~noise, the approach fails for most classes of statistics. Notice that the aforementioned limitations of the multiplicative periodogram bootstrap are common to other frequency domain bootstrap methods which generate independent pseudo-periodogram ordinates. The local periodogram bootstrap introduced by Paparoditis and Politis (1999) is such an example.\\
Thus, for a frequency domain bootstrap procedure to be successful for a wider class of statistics and/or a wider class of stationary processes, it  has to take into account the dependence structure of the ordinary periodogram at different frequencies.
Áttempts in this direction are the approach proposed by Dahlhaus and Janas (1994), the autoregressive-aided periodogram bootstrap by Kreiss and Paparoditis (2003), and the hybrid wild bootstrap, cf. Kreiss and Paparoditis (2012).
Beyond sample mean and nonparametric spectral density estimation, however, the second  approach only works for linear stochastic processes which obey a finite or infinite order autoregressive structure driven by i.i.d.~innovations.
Although the idea behind the hybrid wild bootstrap is different,
and this approach extends the validity of the frequency domain bootstrap to a wider class of statistics compared to the multiplicative periodogram bootstrap, its limitation lies, as for the approach proposed by Janas and Dahlhaus (1994),  in the fact that its applicability is also restricted to linear processes.\\
The above discussion demonstrates that a frequency domain bootstrap procedure which is valid for a wide range of stationary stochastic processes and for a rich class of periodogram-based statistics is missing. This paper attempts to fill this gap. We propose a new  bootstrap approach which is based on bootstrapping periodograms of subsamples of the time series at hand   and which considerably extends the range of validity of frequency domain bootstrap methods. The method    is consistent for a much wider range of stationary processes satisfying very weak dependence conditions. The new approach defines  subsamples  of ''frequency domain residuals'' which are obtained   by appropriately rescaling
 periodograms calculated at the Fourier frequencies of subsamples of the observed time series.  These residuals together with a consistent estimator of the spectral density are used to generate subsamples of  pseudo periodograms which mimic correctly  the weak dependence structure of the periodogram. Aggregating  such  subsampled pseudo periodograms appropriately,
  leads to  a frequency domain bootstrap approach which can be used  to approximate the distribution of the corresponding  statistic of interest.
 We establish consistency of the new frequency domain  bootstrap procedure for a general class of stationary time series, ranging clearly beyond the linear process class and for general periodogram-based statistics known as spectral means and ratio statistics. Spectral means and ratio statistics include several of the commonly used statistics in time series analysis, like sample autocovariances or sample autocorrelations, as special cases. The idea underlying our approach is somehow related  to that  of convolved subsampling which has been recently and independently considered in a different context than ours by Tewes et al.~(2017).
 Furthermore,  and for the case of spectral means, we show how the new procedure can be used   to  benefit from the  advantages of classical procedures to bootstrap the periodogram and at the same time to overcome  their limitations. In particular, the modification of the standard procedures proposed,  uses existing frequency domain methods to replicate the dominant part of the distribution of the statistic of interest and uses the  new concept of convolved bootstrapped periodograms of subsamples to correct for those parts of the distribution which are due to the weak dependence of the periodogram ordinates and which cannot be mimicked by classical procedures.

 \

The paper is organized as follows. Section \ref{sec_univariate} reviews some asymptotic results concerning the class of spectral means and ratio statistics and clarifies the limitations in approximating the distribution of such statistics by frequency domain bootstrap methods which generate independent periodogram ordinates. Section \ref{sec_hybprocedure} introduces the new frequency domain  bootstrap procedure while Section \ref{sec_validity} establishes its asymptotic validity for the entire class of spectral means and ratio statistics and for a very wide class of stationary processes. Section \ref{sec_numresults} discusses the issue of selecting the bootstrap parameters in practice, presents some simulations demonstrating the finite sample performance of the new  bootstrap procedure. Finally, all technical proofs are deferred to the Appendix of the paper.

\

\section{Spectral Means and Ratio Statistics}\label{sec_univariate}

We consider a weakly stationary real-valued stochastic process $ (X_t)_{t \in \Z} $ with mean zero,
absolutely summable autocovariance function $ \gamma $ and spectral density $ f: [-\pi,\pi] $ $  \rightarrow [0,\infty) $. The periodogram of a sample $ X_1,\ldots,X_n $ from this process is defined according to \eqref{periodogram}. While the periodogram is well-known to be an inconsistent estimator of the spectral density $ f(\lambda) $, integrated periodograms form an important class of estimators which are consistent under suitable regularity conditions. For some integrable function $ \varphi: [-\pi,\pi] \rightarrow \R $ the integrated periodogram is defined as
\bee
M(\varphi,I_n)=\int_{-\pi}^{\pi} \varphi(\lambda) I_n(\lambda) \, d\lambda,
\eee
which is an estimator for the so-called \emph{spectral mean} $ M(\varphi,f) $.
A further interesting class of statistics is obtained by scaling a spectral mean statistic by the quantity $ M(1,I_n)$. In particular, this class of statistics, also known as \emph{ratio statistics}, is defined by
\bee
R(\varphi,I_n) = \frac{M(\varphi,I_n)}{M(1,I_n)}\,.
\eee
For practical calculations the integral in $ M(\varphi,I_n) $ is commonly  replaced by a Riemann sum. This is usually done using the Fourier frequencies based on sample size $ n $ which are given by $ \lambda_{j,n}=2\pi j/n $, for $ j \in \mathcal{G}(n) $, with
\be
\mathcal{G}(n) &:=& \{ j \in \Z: \; 1\leq |j| \leq [n/2] \}. \label{defGn}
\ee
The approximation of $ M(\varphi,I_n) $, and respectively of $R(\varphi,I_n)$, via Riemann sum is then given by
\bee
M_{\mathcal{G}(n)}(\varphi,I_n)= \frac{2\pi}{n} \sum_{j \in \mathcal{G}(n)} \varphi(\lambda_{j,n}) I_n(\lambda_{j,n})
\eee
and
\bee
R_{\mathcal{G}(n)}(\varphi,I_n)= \frac{\sum_{j \in \mathcal{G}(n)} \varphi(\lambda_{j,n}) I_n(\lambda_{j,n})}{\sum_{j \in \mathcal{G}(n)} I_n(\lambda_{j,n})} .
\eee

The construction of $ \mathcal{G}(n) $ is motivated by the fact that integrals over $ [0,\pi] $ are usually approximated by Riemann sums at frequencies $ \lambda_{j,n}=2\pi j/n $, for $ j=1,\ldots[n/2] $. Since the periodogram $ I_n $ (and in many applications the function $ \varphi $ as well) is axis-symmetric, it will be useful to employ a Riemann sum over $ [-\pi,\pi] $ where for each used frequency $ \lambda_{j,n} $ the corresponding negative frequency $ -\lambda_{j,n} $ is also used. This is ensured by using $ \lambda_{j,n} $ with $ j \in\mathcal{G}(n) $ instead of $ j=-[n/2]+1,\ldots,[n/2] $.\\
Different statistics commonly used in time series analysis belong to the class of spectral means or ratio statistics. We give some examples.

\begin{example} $ \left.\right. $\\
\begin{enumerate}
\item[(i)]
The autocovariance $ \gamma(h) $ of $ (X_t) $ at lag $ 0 \leq h < n $ can be estimated by the empirical autocovariance $ \widehat{\gamma}(h)=n^{-1} \sum_{t=1}^{n-h} X_t X_{t+h} $ which is an integrated periodogram statistic. This is due to the fact that choosing $ \varphi(\lambda)=\cos(h\lambda) $ it follows from straightforward calculations that $ \widehat{\gamma}(h)=M(\varphi,I_n) $ as well as $ \gamma(h)=M(\varphi,f) $.
\item[(ii)] The spectral distribution function evaluated at point $ x\in(0,\pi] $ is defined as $ \int_{0}^x f(\lambda) \, d\lambda=M(\varphi,f) $ for $\varphi(\lambda)=\mathds{1}_{(0,x]}(\lambda) $. The corresponding integrated periodogram estimator is given by
\begin{eqnarray*}
M(\varphi,I_n)=\int^x_{0} I_n(\lambda) \, d\lambda.
\end{eqnarray*}
\item[(iii)] The autocorrelation $\rho(h)=\gamma(h)/\gamma(0)$  of   $ (X_t) $ at lag $ 0 \leq h < n $ can be estimated by the empirical
autocorrelation $ \widehat{\rho}(h)=\widehat{\gamma}(h)/\widehat{\gamma}(0)$ which in view of (i) is a  ratio statistic, that is
$ \widehat{\rho}(h)=R(\cos(h\cdot),I_n)$.
\end{enumerate} $ \hfill \square $
\end{example}

We summarize the assumptions imposed on the process $ (X_t)_{t \in \Z} $ and the function $ \varphi $:

\begin{assumption} \label{assu_1} $ \left.\right. $\\
\begin{enumerate}
\item[(i)]
Assumptions on $ (X_t)_{t \in \Z} $: Let $ (X_t)_{t \in \Z} $ be a strictly stationary, real-valued stochastic process with finite eighth moments, mean zero,  autocovariance function $ \gamma: \Z \rightarrow \R $ fulfilling $ \sum_{h\in \Z} (1+|h|)|\gamma(h)| <\infty $ and spectral density $ f $ satisfying $ \inf_{\lambda\in[0,\pi]}f(\lambda) >0$.
 Furthermore, let the fourth order joint cumulants of the process fulfil
\bee
\sum_{h_1,h_2,h_3\in \Z} (1+|h_1|+|h_2|+|h_3|) \, |\textrm{cum}(X_0,X_{h_1},X_{h_2},X_{h_3})|<\infty \,,
\eee
and the eighth order cumulants be absolutely summable, that is
\bee
\sum_{h_1,\ldots,h_7\in \Z} |\textrm{cum}(X_0,X_{h_1},\ldots,X_{h_7})|<\infty \,.
\eee
\item[(ii)]
Assumptions on $ \varphi $: Let $ \varphi: [-\pi,\pi] \rightarrow \R $ be a square--integrable function of bounded variation.
\end{enumerate} $ \hfill \square $
\end{assumption}

Notice that the  summability conditions imposed on the autocovariance function  $\gamma$,  imply boundedness and differentiability of $ f $, as well as boundedness of the derivative  $ f' $ of $f$. Under the conditions of Assumption \ref{assu_1}, and some additional weak dependence conditions, it is known that $ M(\varphi,I_n) $ is a consistent estimator for $ M(\varphi,f) $ and that the following central limit theorem holds true:
\be
L_n=\sqrt{n} \big(M(\varphi,I_n)-M(\varphi,f)\big) \stackrel{d}{\longrightarrow} \mathcal{N}(0,\tau^2), \label{CLT}
\ee
with $ \tau^2=\tau_1^2+\tau_2 $, where
\be
\tau_1^2 &=& 2\pi \, \int_{-\pi}^{\pi} \varphi(\lambda) \big( \varphi(\lambda) + \varphi(-\lambda) \big) f(\lambda)^2 \, d\lambda\,, \label{limit_var1}\\
\tau_2 &=& 2\pi \, \int_{-\pi}^{\pi} \int_{-\pi}^{\pi} \varphi(\lambda_1) \varphi(\lambda_2) f_4(\lambda_1,\lambda_2,-\lambda_2) \, d\lambda_1 \, d\lambda_2 \,, \label{limit_var2}
\ee
and where
\bee
f_4(\lambda,\mu,\eta)= \frac{1}{(2\pi)^3} \sum_{h_1,h_2,h_3\in \Z} \textrm{cum}(X_0,X_{h_1},X_{h_2},X_{h_3}) \, e^{-i(h_1\lambda+ h_2\mu + h_3\eta)}
\eee
is the fourth order cumulant spectral density, cf. \citeasnoun{Rosenblatt85}, Chapter III, Corollary 2.

\begin{noindent}
\begin{remark} \label{rem_simplifications}
$ (i) $ In the literature the function $ \varphi $ is sometimes assumed to be even, i.e. $ \varphi(\lambda)=\varphi(-\lambda) $ for all $ \lambda $. In this case the first part $ \tau_1^2 $ of the limiting variance takes the form
\bee
4\pi \, \int_{-\pi}^{\pi} \varphi(\lambda)^2 f(\lambda)^2 \, d\lambda = 8\pi \, \int_{0}^{\pi} \varphi(\lambda)^2 f(\lambda)^2 \, d\lambda \,.
\eee
However, we allow for general functions $ \varphi $ since this restriction to even functions is not necessary.\\
$(ii)$  Approaches to directly estimate the integral involving  the fourth order cumulant spectral density and which can potentially  be used  to estimate the variance $\tau^2$ have been proposed by some authors; see Tanuguchi (1982), Keenan (1987) and Chiu (1988). However, the empirical performance of such estimators is unknown. Alternatively, Shao (2009) proposed an approach to construct confidence intervals for $ M(\varphi,f)$ based on self normalization
which bypasses the problem of estimating  the variance term $\tau^2$. \\
$ (iii) $ The second summand $ \tau_2 $ of the limiting variance $ \tau^2 $ simplifies if the process $ (X_t) $ is a linear process, that is, if $ (X_t) $ admits the representation \eqref{linearproc} for some square-summable sequence of coefficients $ (a_j)_{j \in \Z} $ and some i.i.d.~white noise process $ (\varepsilon_t)_{t \in \Z} $ with finite fourth moments. Denoting $ \sigma^2:=E(\varepsilon_1^2) $ and $ \eta:=E(\varepsilon_1^4)/\sigma^4 $, it then follows by a standard calculation the decomposition $ 2\pi \, f_4(\lambda_1,\lambda_2,-\lambda_2) = (\eta -3)f(\lambda_1) f(\lambda_2) $. In this case, the second summand $ \tau_2 $ of the limiting variance from \eqref{CLT} can be written as
\be
\tau_{2,\textrm{lin}}= (\eta -3) \, \left(\int_{-\pi}^{\pi} \varphi(\lambda) f(\lambda) \, d\lambda \right)^2. \label{limit_var_lin}
\ee
$ \hfill \square $
\end{remark}
\end{noindent}

Regarding the class of ratio statistics the situation is somehow different. Notice first that
\begin{align*}
L_{n,R}& :=\sqrt{n} \big(R(\varphi,I_n)-R(\varphi,f)\big)\\
 & = \frac{1}{M(1,I_n) M(1,f)} \sqrt{n}\int_{-\pi}^{\pi} w(\lambda) I_n(\lambda) \, d\lambda,
\end{align*}
where
\[  w(\lambda) = \varphi(\lambda) \int_{-\pi}^{\pi}f(\alpha) \, d\alpha - \int_{-\pi}^{\pi}\varphi(\alpha)f(\alpha) \, d\alpha.\]
In view of \eqref{CLT} and the fact that  $M(1,I_n)\stackrel{P}{\rightarrow} M(1,f)$ we then have that
\begin{equation} \label{CLT-Ratio}  L_{n,R} \stackrel{d}{\longrightarrow} \mathcal{N}(0,\tau^2_R),
\end{equation}
where $\tau^2_R=\tau^2_{1,R} + \tau_{2,R}$ with
\be
\tau^2_{1,R} & = &\frac{2\pi}{M^4(1,f)}\int_{-\pi}^{\pi}w(\lambda)\big(w(\lambda)+w(-\lambda) \big)f(\lambda)^2 \, d\lambda,\\
\tau_{2,R}&=& \frac{2\pi}{M^4(1,f)}\int_{-\pi}^{\pi}\int_{-\pi}^{\pi}w(\lambda_1)w(\lambda_2) f_4(\lambda_1,\lambda_2,-\lambda_2) \, d\lambda_1d\lambda_2\,. \label{limit_var2R}
\ee
It  can be easily verified that  $ \int_{-\pi}^{\pi}w(\lambda)f(\lambda) \, d\lambda=0$. This implies that for the class of linear processes considered in Remark 2.2 $(iii)$ and by the same arguments to  those used there, we get that $ \tau_{2,R}=0$. That is, the variance of the limiting Gaussian distribution (\ref{CLT-Ratio}) simplifies for linear processes to $ \tau^2_R=\tau^2_{1,R}$ and it is, therefore, not affected by the fourth order structure of the process. Note that this simplification for ratio statistics holds no longer true when considering nonlinear processes.

\section{The frequency  domain bootstrap procedure} \label{sec_hybprocedure}

A common way to approximate the distribution of $ L_n=\sqrt{n} (M(\varphi,I_n)-M(\varphi,f)) $ based on a given sample $ X_1,\ldots,X_n $ is to use the following multiplicative periodogram bootstrap (MPB)  approach: Approximate $ \mathcal{L}(L_n) $ by the distribution of
\be
V_{n}^*=\sqrt{n} \left( \frac{2\pi}{n} \sum_{j \in \mathcal{G}(n)} \varphi(\lambda_{j,n}) \left( T^*(\lambda_{j,n}) - \widehat{f}_n(\lambda_{j,n}) \right) \right), \label{Ln1star}
\ee
where $ \widehat{f}_n $ is a consistent (e.g.~kernel-type) estimator of $ f $ based on $ X_1\ldots,X_n $, and
\begin{equation} \label{eq.MPB-eq1}
T^*(\lambda_{j,n}):=\widehat{f}_n(\lambda_{j,n}) \, U_j^*,
\end{equation}
where the bootstrap random variables $ (U_j^*)_{j \in \mathcal{G}(n)} $ are constructed in the following way, cf. \citeasnoun{FraHar92}: For $ j > 0 $, let $ U_1^*,\ldots,U_{[n/2]}^* $ be i.i.d. with a discrete uniform distribution on the set of  rescaled ''frequency domain residuals''
\[  \widehat{U}_j= \widetilde{U}_j/ \overline{U}_n, \ \ j=1, 2, \ldots, [n/2],\]
where $\widetilde{U}_j=I_n(\lambda_{j,n})/\widehat{f}_n(\lambda_{j,n})$ and $ \overline{U}_n=[n/2]^{-1}\sum_{j=1}^{[n/2]} \widetilde{U}_j$.
Then, for $ j<0 $, set $ U_j^*:=U_{-j}^* $. The construction with $ U_j^*=U_{-j}^* $ ensures that $ T^*(\cdot) $ is an even function which preserves an important property of the periodogram $ I_n(\cdot) $. To see the motivation for approach \eqref{Ln1star}, notice that $ V_{n}^* $ is a Riemann approximation for $ \sqrt{n} (M(\varphi,T^*)-M(\varphi,\widehat{f}_n)) $. Moreover, the bootstrap random variables $ T^*(\lambda_{j,n}) $ are supposed to mimic the behavior of the peridogram ordinates $ I_n(\lambda_{j,n}) $. Notice that  based on the fact that $ I_n(\lambda)/f(\lambda) $ has an asymptotic standard exponential distribution for $ \lambda \in (-\pi,\pi)\setminus\{0\} $, an alternative approach would be to  generate the $ U^\ast_j$'s as  i.i.d.~exponentially distributed random variables with parameter $ 1 $. Since the asymptotic behavior of  MPB procedure is identical for both approaches used  to generate the pseudo innovations $U^*_j$,  see Dahlhaus and Janas (1996), for simplicity we concentrate in the following  on the  approach generating the $U^*_j$'s as standard exponential  distributed random variables.

\

Under mild conditions $ V_{n}^* $ has a limiting normal distribution with mean zero and variance $ \tau_1^2 $ as defined in \eqref{limit_var1}. The proof is given in Proposition \ref{prop1}. For the special case of linear processes this asymptotic result for integrated periodograms was obtained in \citeasnoun{DahlhausJanas96}. Moreover, \citeasnoun{ShaoWu07} obtain a similar result for smoothed periodogram spectral density estimators and a general class of nonlinear processes. Hence, for a quite general class of stationary processes, the bootstrap approach \eqref{Ln1star} correctly captures the first summand $ \tau_1^2 $ but fails to mimic the second summand $ \tau_2 $ of the limiting variance $ \tau^2 $ from \eqref{CLT}. Consequently, the approach asymptotically only works in those special cases where $ \tau_2 $ vanishes. For the class of linear processes -- where $ \tau_2 $ takes the form $ \tau_{2,\textrm{lin}} $ from \eqref{limit_var_lin} -- this is the case, for example, if the innovations are Gaussian which implies $ \eta=3 $. Another example for linear processes occurs when one is estimating a spectral mean identical to zero, i.e $ \int_{-\pi}^{\pi} \varphi(\lambda) f(\lambda) \, d\lambda=0 $, which is the case for ratio statistics, cf. the discussion in the previous section. But except for these special cases the approach via $ V_{n}^* $ fails in general, especially if the underlying process $ (X_t) $ is not a linear process driven by   i.i.d.~innovations.

\

In the following, we propose an alternative bootstrap quantity,  denoted by $ L_{n}^* $, that is capable of mimicking the distribution of $ L_n$ including
the entire limiting variance of $L_n$ as given in \eqref{CLT}. The modified approach is based on   periodograms calculated on subsamples. More precisely,  the periodogram of each subsample  of length $ b<n $ of  the original time series  is first
 calculated.  These  periodograms  are then appropriately rescaled leading to  a set of subsamples of  ``frequency domain residuals".
 A number $ k=[n/b]$  of   randomly selected subsamples of such residuals  is  then chosen which multiplied  with the spectral density estimator $\widehat{f}_n$ at the appropriate set of Fourier frequencies, leads to the generation of  subsamples of  pseudo periodograms that correctly imitate the weak dependence structure of the ordinary  periodogram.
The convolution of these subsampled pseudo  periodograms is then used for the construction of a bootstrap quantity $ L_{n}^*$ which is used to approximate the distribution of  $L_n$ of interest.  We
show in Theorem \ref{theorem1} that  this new approach is consistent under very mild conditions on the underlying stochastic process. Furthermore, the same approach  can be used in a  hybrid type bootstrap procedure, which is  denoted by $ \widetilde{V}_{n}^* $, and which is composed from both $ V_{n}^* $ and $ L_{n}^* $ and corrects for the  drawbacks of the MPB procedure based solely on $V^*_n$.

\

We will now first state the proposed basic bootstrap algorithm and discuss some of its properties
in a series of remarks.

\

\noindent{\bf Convolved Bootstrapped Periodograms of  Subsamples}\\
\begin{enumerate}
\item  Let $\widehat f_n$ be some consistent spectral density estimator, for instance, the estimator used in (\ref{eq.MPB-eq1}).
Choose a block length $ b<n $, assume that $ n=bk $ with $ b,k \in \N $. For $t=1,2, \ldots, N$, with $N:=n-b+1$, let
\[ I_{t,b}(\lambda_{j,b})=\frac{1}{2\pi b} \left| \, \sum_{s=1}^{b} X_{t+s-1} \, e^{-i\lambda_{j,b} s} \, \right|^2,\]
 where $\lambda_{j,b} $ are Fourier frequencies associated with subsample series of length $ b $, i.e. $ \lambda_{j,b}=2\pi j/b $, with $ j\in \mathcal{G}(b) $, cf. \eqref{defGn} for the definition of $ \mathcal{G}(\cdot) $.
\item
Define the rescaled frequency domain residuals of the subsampled periodogram $  I_{t,b}(\lambda_{j,b})$ as,
\[ U_{t,b}(\lambda_{j,b}) = \frac{ I_{t,b}(\lambda_{j,b})}{\widetilde{f}_b(\lambda_{j,b})}, \ j=1,2, \ldots, [b/2],\]
where  $ \widetilde{f}_b(\lambda_{j,b})=N^{-1} \sum_{t=1}^{N} I_{t,b}(\lambda_{j,b})$.
\item
Generate random variables $ i_1^*,\ldots,i_k^* $ i.i.d.~with a discrete uniform distribution on the set $ \{1,\ldots,N \} $.
For  $  l=1,2, \ldots, k$, define
\bee
I_b^{(l)}(\lambda_{j,b})= \widehat f_n(\lambda_{j,b}) \cdot U_{i_l^*,b}(\lambda_{j,b}),
\eee
and  let
\bee
I_{j,b}^*=\frac{1}{k} \sum_{l=1}^{k} I_b^{(l)}(\lambda_{j,b}) \,.
\eee
\item
Approximate the distribution of $ L_n=\sqrt{n} (M(\varphi,I_n)-M(\varphi,f)) $ by the distribution of the  bootstrap quantity
\be
L_n^*
 &:=& \sqrt{n} \, \frac{2\pi}{b} \sum_{j \in \mathcal{G}(b)} \varphi(\lambda_{j,b}) \left( I_{j,b}^* - \widehat{f}_n(\lambda_{j,b}) \right) \,. \label{V2star}
\ee
Furthermore, approximate the distribution of  $ L_{n,R}=\sqrt{n} (R(\varphi,I_n)-R(\varphi,f)) $ by that of
\be
L_{n,R}^* &:=& \sqrt{n} \, \left(  \frac{\sum_{j\in \mathcal{G}(b)} \varphi(\lambda_{j,b}) I^*_{j,b}}{\sum_{j\in \mathcal{G}(b)} I^*_{j,b}} - \frac{\sum_{j\in \mathcal{G}(b)} \varphi(\lambda_{j,b}) \widehat{f}_n(\lambda_{j,b})}{\sum_{j\in \mathcal{G}(b)} \widehat{f}_n(\lambda_{j,b})} \right) \,. \label{V2star-Ratio}
\ee
\end{enumerate}
$ \hfill \square $

\
Some remarks are in order.
\

\

\begin{remark} \label{remark0a}\
\begin{enumerate}
\item[]  $(i)$ The above   approach differs from subsampling since we do not calculate  $L_n$ (respectively $ L_{n,R}$) solely on a subsample of the observed time series. Similarly,  the pseudo periodograms   $ I^{(l)}_b(\lambda_{j,b})$, $l=1,2, \ldots, k$, as well as
$L_n$ (respectively $L_{n,R}$),  are  not  calculated on $k$ randomly selected subsamples of the observed time series.  In fact, our procedure generates new pseudo periodograms of subsamples   $ I^{(l)}_b(\lambda_{j,b})$, $l=1,2, \ldots, k$,   using the spectral density estimator $ \widehat{f}_n$  based on the entire time series  $ X_1, X_2, \ldots, X_n$ and the set of appropriately defined subsamples of frequency domain residuals $ \{U_{t,b}(\lambda_{j,b}), j=1,2, \ldots, [b/2]\}$ and $ t=1,2, \ldots, N$.
\item[] $(ii)$   Similar to  the  MPB, the rescaling in (2) ensures that  $E^*(U_{i_l^*,b}(\lambda_{j,b}) )=1$, i.e.,
$E^*( I^{(l)}_b(\lambda_{j,b}))=\widehat{f}_n(\lambda_{j,b})$, which avoids  an unnecessary bias at the resampling step.
\item[] $(iii)$   Notice that $L_n^*$ can be written as $L_n^*=k^{-1}\sum_{l=1}^k L_{l,n}^*$, where
\[ L^*_{l,n}= \sqrt{n}\frac{2\pi}{b} \sum_{j\in {\mathcal G}(b)} \varphi(\lambda_{j,b})\widehat{f}_n(\lambda_{j,b})\big(U_{i_l^*,b}(\lambda_{j,b})-1\big).\]
Comparing the above expression with that of the MPB    given by
\[ V^*_n=\sqrt{n}\frac{2\pi}{n} \sum_{j\in {\mathcal G}(n)} \varphi(\lambda_{j,n})\widehat{f}_n(\lambda_{j,n})\big(U^*_{j}-1\big),\]
clarifies  that, apart from the different number of observations  on which $L^*_{l,n}$   and $V^*_n $ are based, the essential difference lies in  the way the pseudo innovations in the two approaches  are generated. In particular, while the $ U^*_j $'s  are independent  the  pseudo random variables $ U_{i_l^*,b}(\lambda_{j,b})$ are \emph{not}. In fact, due to resampling the entire subsample  of  frequency domain residuals
$ U_{i_l^*,b}(\lambda_{j,b})$,
in generating   the  subsampled pseudo periodogram
$ I^{(l)}_b(\lambda_{j,b})$, $j=1,2, \ldots, [b/2]$,  the weak dependence structure of the periodogram within the subsamples is preserved.

\item[] $(iv)$ Note that the rescaling  quantity $ \widetilde f_b $ used in (2) is actually itself a spectral density estimator which is based on averaging periodograms calculated over subsamples. Such  estimators have been thoroughly investigated by many authors in the literature;  Bartlett  (1948), (1950), Welch (1967); see  also
\citeasnoun{Dahlhaus85}.  We will make use of some of the  results obtained for this estimator later on.
This implies that we could also set  $ \widehat{f}_n=\widetilde{f}_b$  in (3) of the bootstrap procedure
in order to  generate the subsamples of pseudo periodograms $ I^{(l)}_b(\lambda_{j,b})$.   However, allowing for the use of   any consistent spectral density estimator $ \widehat f_n $  makes  the bootstrap approach much more flexible. Depending on the data sample at hand, either appropriate parametric or nonparametric estimators may be chosen from case to case.
\end{enumerate}
\end{remark} $ \hfill \square $

%

\

As mentioned in the Introduction the new approach to bootstrap the statistics $L_n$ (respectively $L_{n,R}$) can also be used in a hybrid type procedure  in the case of spectral maens in order to  correct for the shortcomings  of standard  frequency domain bootstrap methods, like for instance  of the MPB procedure. In the following we formulate such a hybrid frequency domain bootstrap approach which, as we will see in Theorem~\ref{theorem2}, makes the MPB procedure consistent for the entire class of spectral means and ratio statistics.

\

\noindent{\bf A hybrid periodogram bootstrap (spectral means)}\\
\begin{enumerate}
\item Let $L^*_n$ be generated according to \eqref{V2star} -- that is, as in the  bootstrap procedure  generating subsamples of pseudo periodograms -- and let $V_n^*$ be defined as in (\ref{Ln1star}), where the i.i.d. random variables $U^*_j$, $j=1,2, \dots, [n/2]$,  and $ i_1^*, i_2^*,\ldots, i_k^*$ are independent.
\item
Approximate the distribution of $ L_n=\sqrt{n} (M(\varphi,I_n)-M(\varphi,f)) $ by the empirical distribution of the rescaled bootstrap quantity
\bee
\widetilde{V}_n^*:=\frac{\widehat \tau}{\widehat \tau_1} V_{n}^* \,,
\eee
with the standardization quantity $ \widehat \tau_1^2:={\rm Var}^*(V_{n}^*) $, and with a variance correction factor $ \widehat \tau^2:=
{\rm Var}^*(L_n^*+V_{n}^*)-c_n $, where
\bee
c_n= \frac{4\pi^2}{b} \sum_{j \in \mathcal{G}(b)} \varphi(\lambda_{j,b})\, \big( \varphi(\lambda_{j,b}) + \varphi(-\lambda_{j,b}) \big)\widehat{f}_n(\lambda_{j,n})^2\left( \frac{1}{N} \sum_{t=1}^{N} \frac{I_{t,b}(\lambda_{j,b})^2}{\widetilde{f}_b(\lambda_{j,b})^2} - 1 \right) \,.
\eee
$ \hfill \square $
\end{enumerate}

\begin{remark}  \label{remark1a}
The idea of the above hybrid procedure is to base bootstrap approach $ \widetilde{V}_n^* $ on $ V_{n}^* $ and to perform a variance correction to achieve consistency for general stationary processes. As argued before, $ V_{n}^* $ has a limiting normal distribution with mean zero and variance $ \tau_1^2 $. It is not able to mimic the fourth order cumulant term $ \tau_2 $. As can be seen by inspecting the proof of Proposition \ref{prop1}, this is due to the fact that the bootstrap ordinates $ T^*(\lambda_{j,n}) $ and $ T^*(\lambda_{k,n}) $ are (conditionally) independent whenever $ |j|\neq |k| $, while periodogram ordinates at different frequencies are in general dependent. To be precise, it is well--known that for different \emph{fixed} frequencies periodogram ordinates are only \emph{asymptotically} independent but for finite sample size correlated. When using Fourier frequencies $ 2\pi j/n $ instead of fixed frequencies, this effect of interdependence does not vanish asymptotically but shows up in form of summand $ \tau_2 $ in the limiting variance. The other summand $ \tau_1^2 $ in turn evolves from the sum of the variances of the periodogram ordinates at different frequencies without taking dependencies into account; and this contribution to the limiting variance is mimicked correctly by $ V_{n}^* $.\\
In approach $ \widetilde{V}_n^*=(\widehat \tau/\widehat \tau_1) V_{n}^* $ we therefore use $ L_{n}^* $ solely to mimic $ \tau_2 $. The factor $ 1/\widehat \tau_1 $ standardizes $ V_{n}^* $ and the factor $ \widehat \tau $ establishes the appropriate variance. We have $ \widehat \tau^2={\rm Var}^*(L^*_n+V_{n}^*)-c_n $. Since the random variables $ V_{n}^* $ and $ L_{n}^* $ are conditionally independent,
$ L_n^* + V^*_n $ is asymptotically normal with mean zero and variance $ 2\tau_1^2+\tau_2 $. Hence, the asymptotic variance has to be reduced by $ \tau_1^2 $ to achieve consistency. To be precise, the contribution of $ L_{n}^* $ to $ 2\tau_1^2 $ has to be removed since, then, $ V_{n}^* $ contributes the $ \tau_1^2 $ part and $ L_{n}^* $ the necessary correction for $ \tau_2 $. This is done in step $ (2) $ of the hybrid bootstrap algorithm: Note that by the proof of Theorem \ref{theorem1} $(i)$ $ c_n \rightarrow \tau_1^2 $ in probability, as $ n \rightarrow \infty $, so that $ \widehat \tau^2 \rightarrow \tau^2 $ in probability. On the one hand, this implies asymptotic consistency of the bootstrap approach for general stationary processes. On the other hand, $ \widehat \tau^2 $ is also the appropriate variance correction for finite sample sizes because $ c_n $ represents exactly the contribution of $ L_{n}^* $ to $ \tau_1^2 $, as can be seen by the proof of Theorem \ref{theorem1} $(i)$. $ \hfill \square $
\end{remark}

\begin{remark}
The modification of the multiplicative periodogram bootstrap proposed leads to  a potential   advantage  of this procedure compared to the one based solely on
convolving bootstrapped  periodograms of subsamples. To elaborate, consider for instance   the case of  spectral means and recall that in the modification proposed, the estimator of the first part $ \tau_1^2$ of the variance is  delivered by the multiplicative periodogram bootstrap
while the procedure based on convolved bootstrapped periodograms is only  used in order to estimate the second part $ \tau_2$ of the variance. This, however, implies  that
the  hybrid bootstrap estimator of the distribution of the statistic of interest becomes  less sensitive with respect to the  choice of the subsampling parameter  $b$,  compared to  the procedure based exclusively on convolved bootstraped  periodograms. This aspect will be  demonstrated  in  Section  5.
\end{remark}

The next remark deals with the question how the above hybrid bootstrap procedure can be implemented in practice.

\begin{remark} \label{remark1a}
The bootstrap variances $ {\rm Var}^*(L^*_n+V_{n}^*) $ and $ {\rm Var}^*(V_{n}^*) $ used in step $ (2) $ of the hybrid bootstrap approach are not readily available when implementing the algorithm in practice. Hence, they have to be replaced by estimators. One possibility is via Monte Carlo: Repeat step $ (1) $ multiple times to generate $ M $ replications\\ $ V_{n}^*(1),\ldots,V_{n}^*(M) $ of $ V_{n}^* $. Then calculate the variance estimator
\begin{align*}
\widetilde \tau_{1}^2 = \frac{1}{M-1} \sum_{j=1}^{M} \left( V_{n}^*(j)-\overline{V}_{n}^* \right)^2 \,,
\end{align*}
where $ \overline{V}_{n}^*=M^{-1} \sum_{j=1}^{M} V_{n}^*(j) $. Proceed analogously to replace $ {\rm Var}^*(L^*_n+V_{n}^*) $ in $ \widehat \tau^2 $ and denote the estimated version by $ \widetilde \tau^2 $. Finally, obtain rescaled bootstrap replications $ \widetilde{V}_n^*(1),\ldots,\widetilde{V}_n^*(M) $, via
\bee
\widetilde{V}_n^*(j)=\frac{\widetilde \tau}{\widetilde \tau_{1}} V_{n}^*(j), \quad \quad j=1,\ldots,M \,.
\eee
$ \hfill \square $
\end{remark}

We conclude this section by a modification of the hybrid bootstrap proposal  which is appropriate to approximate the distribution of ratio statistics.
This modification is necessary due to the normalizing term $ M(1,I_n)M(1,f)$; see the definition of $L_{n,R}$.

\

\noindent{\bf  A hybrid periodogram bootstrap (ratio statistics)}\\
\begin{enumerate}
\item
Let
\be \label{eq.diffVs}
V^*_{1,n} &=&\frac{2\pi}{\sqrt{n}} \sum_{j \in \mathcal{G}(n)}  \widehat{w}(\lambda_{j,n}) \, T^*(\lambda_{j,n})\,, \nonumber\\
V_{2,n}^* &=& \sqrt{n} \, \frac{2\pi}{b} \sum_{j \in \mathcal{G}(b)} \widetilde{w}(\lambda_{j,b}) \,I_{j,b}^* \,, \\
V_{n,R}^* &=& \frac{1}{D^*_n} V^*_{1,n} \,,  \nonumber
\ee
where
\[
D^*_n =\frac{2\pi}{n} \sum_{j\in\mathcal{G}(n)} T^\ast(\lambda_{j,n}) \, \frac{2\pi}{n} \sum_{j\in\mathcal{G}(n)}\widehat{f}_n(\lambda_{j,n}) \,, \]
\be
\widehat{w}(\lambda)= \varphi(\lambda) \frac{2\pi}{n}\sum_{j\in \mathcal{G}(n)}\widehat{f}_n(\lambda_{j,n}) - \frac{2\pi}{n} \sum_{j\in \mathcal{G}(n)}\varphi(\lambda_{j,n})\widehat{f}_n(\lambda_{j,n}) \,, \label{what}
\ee
and
\be
\widetilde{w}(\lambda)= \varphi(\lambda) \frac{2\pi}{b}\sum_{j\in \mathcal{G}(b)}\widehat{f}_n(\lambda_{j,b}) - \frac{2\pi}{b} \sum_{j\in \mathcal{G}(b)}\varphi(\lambda_{j,b})\widehat{f}_n(\lambda_{j,b}). \label{wtilde}
\ee
Finally,  let $V^*_{3,n}=V_{1,n}^* + V_{2,n}^*$.
\item  Approximate the distribution of $ L_{n,R}=\sqrt{n} (R(\varphi,I_n)-R(\varphi,f)) $ by the distribution of  $ \widetilde{V}_{n,R}^* $, where
\bee
\widetilde{V}_{n,R}^*=\frac{\widehat \sigma_R}{\widehat{\sigma}_{1,R}} V_{n,R}^*\,,
\eee
where $ \widehat{\sigma}^2_{1,R} = {\rm Var}^*(V_{1,n}^*)$, $ \widehat{\sigma}^2_R={\rm Var}^*(V_{3,n}^*) -c_{n,R}$,
and
\bee
c_{n,R} = \frac{4\pi^2}{b} \sum_{j \in \mathcal{G}(b)} \widetilde{w}(\lambda_{j,b}) \big( \widetilde{w}(\lambda_{j,b}) + \widetilde{w}(-\lambda_{j,b}) \big) \widehat{f}_n(\lambda_{j,b})^2\left( \frac{1}{N} \sum_{t=1}^{N} \frac{I_{t,b}(\lambda_{j,b})^2}{\widetilde{f}_b(\lambda_{j,b})^2}-1 \right) \,.
\eee
$ \hfill \square $
\end{enumerate}

\

\begin{remark} Notice that $ 2\pi n^{-1}\sum_{j\in \mathcal G (n)} \widehat{w}(\lambda_{j,n} )\widehat{f}_n(\lambda_{j,n})=0$ and
$2\pi b^{-1}\sum_{j\in \mathcal G (b)}$ $  \widetilde{w}(\lambda_{j,b} ) $ $ \widehat{f}_n(\lambda_{j,b})=0$   which makes the centering of $ V_{1,n}^*$ and $ V_{2,n}^*$
in \eqref{eq.diffVs} obsolete.
\end{remark}

\begin{remark} As for spectral means the distribution of $ \widetilde{V}^*_{n,R}$ can be evaluated by Monte Carlo. In particular, we may repeat step $(1)$ multiple times to generate $M$ replications  $ V_{n,R}^*(j)$, $ V^*_{s,n}(j)$, $ s=1,2, 3$, $j=1,2,  \ldots, M$, and  we can then  calculate
$\widetilde{V}_{n,R}^*(j)=\widetilde\sigma_R \cdot V_{n,R}^*(j)\big/  \widetilde \sigma_{1,R} $, $j=1,2, \ldots, M$. Here,
$\widetilde \sigma_R^2=(\widetilde \sigma_n^2 - c_{n,R})$,
 $ \widetilde{\sigma}_n^2 = (M-1)^{-1} \sum_{j=1}^{M} \left(V_{3,n}^*(j)-\overline{V}_{3,n}^* \right)^2 $ and $ \widetilde \sigma_{1,R}^2 = (M-1)^{-1} \sum_{j=1}^{M} \left( V_{1,n}^*(j)-\overline{V}_{1,n}^* \right)^2$. The empirical distribution of the replications
$ \widetilde{V}_{n,R}^*(j)$, $j=1,2, \ldots, M$, can then be used to estimate the  distribution  of $ \widetilde{V}_{n,R}^*$. $ \hfill \square $
\end{remark}

\section{Bootstrap Validity} \label{sec_validity}

In order to establish consistency for the aforementioned bootstrap approaches, we impose the following assumption on the asymptotic behavior of the subsample block length $ b $:

\begin{assumption} \label{assu_bk} $ \left.\right. $\\
\begin{enumerate}
\item[(i)] For block length $ b $ and $ k=n/b $, it holds $ b=b(n) $ and $ k=k(n) $, such that $ b \rightarrow \infty $ and $ k \rightarrow \infty $, as $ n \rightarrow \infty $.
\item[(ii)] It holds $ b^3/n \rightarrow 0 $ and $ \ln(N)/b \rightarrow 0 $, as $ n \rightarrow \infty $.
\end{enumerate} $ \hfill \square $
\end{assumption}
For the bootstrap approaches  considered,  we assume uniform consistency for the spectral density estimator $ \widehat{f}_n $, that is:

\begin{assumption} \label{assu_fnhat}
The spectral density estimator $ \widehat{f}_n $ fulfils
\bee
\sup_{\lambda \in [-\pi,\pi]} \big| \widehat{f}_n(\lambda)-f(\lambda) \big| =o_P(1) \,.
\eee
$ \hfill \square $
\end{assumption}
This is a common and rather weak assumption in the spectral analysis of time series because the rate of convergence is not specified.\\
The upcoming proposition states an asymptotic result for bootstrap approach $ V_{n}^* $. \citeasnoun{DahlhausJanas96} proved this result for the special case of linear processes as given by \eqref{linearproc}. However, since this restriction is not necessary for the bootstrap quantities, we derive the limiting distribution under the aforementioned assumptions for general stationary processes.\\
The following two preliminary results will be useful for the proof of consistency for bootstrap approach $ L_{n}^* $.

\begin{lemma} \label{lemma1}
Under the conditions of Assumption \ref{assu_1} and \ref{assu_bk} $ (i) $ it holds for the Fourier frequencies $ \lambda_{j,b}=2\pi j/b $, $ j \in \mathcal{G}(b) $:
\begin{align*}
(i)& \quad  \sum_{j \in \mathcal{G}(b)} \big| \widetilde{f}_b(\lambda_{j,b}) - EI_{1,b}(\lambda_{j,b}) \big| = \mathcal{O}_P\big( \sqrt{b^3/N} \big),\\
(ii)& \quad \sum_{j,s \in \mathcal{G}(b)} \Big| \frac{1}{N} \sum_{t=1}^{N} I_{t,b}(\lambda_{j,b}) I_{t,b}(\lambda_{s,b}) - E(I_{1,b}(\lambda_{j,b}) I_{1,b}(\lambda_{s,b})) \Big| = \mathcal{O}_P\big( \sqrt{b^5/N} \big).
\end{align*}
\end{lemma}

\

\begin{lemma} \label{lemma2}
Let Assumptions \ref{assu_1} and \ref{assu_bk} as well as assertion \eqref{CLT} hold. Then, for
\be
W_{t,b}:= \frac{2\pi}{\sqrt{b}} \sum_{j \in \mathcal{G}(b)} \varphi(\lambda_{j,b}) \left( I_{t,b}(\lambda_{j,b}) - \widetilde{f}_b(\lambda_{j,b}) \right) \,, \label{MMWtb}
\ee
it holds, as $ n \rightarrow \infty $:
\begin{enumerate}
\item[(i)] $ E(W_{1,b}^2) \rightarrow \tau^2 $, with $ \tau^2 $ as in \eqref{CLT}.
\item[(ii)] $ W_{1,b} \stackrel{d}{\rightarrow} \mathcal{N}(0,\tau^2) $.
\end{enumerate}
\end{lemma}

\

In the following, $ P^* $ will denote conditional probability given the data $ X_1,\ldots,X_n $, and $ E^* $ and $ \textrm{Var}^* $ will denote the corresponding expectations and variances.

\begin{proposition} \label{prop1}
Let Assumptions \ref{assu_1}, \ref{assu_bk} and \ref{assu_fnhat} be fulfilled. Moreover, let $ \Phi $ denote the cdf of the standard normal distribution. Then, with $ \tau_1^2$ as defined in \eqref{limit_var1}, it holds
\begin{enumerate}
\item[(i)] $ {\rm Var}^*(V_{n}^*) \stackrel{P}{\longrightarrow} \tau_1^2,$
\item[(ii)] $ \sup_{x \in \R} | P^*(V_{n}^* \leq x) - \Phi(x/\tau_1) | =o_P(1) $.
\end{enumerate}
\end{proposition}

\

Now, we consider the case of spectral means and ratio statistics and we will show that both bootstrap approaches $ L_n^* $ and $ L_{n,R}^* $   consistently estimate the variance and the distribution of interest.

\

\begin{theorem} \label{theorem1}
Let the assumptions of Proposition~\ref{prop1} as well as assertion \eqref{CLT} be fulfilled. Then, with $ \tau^2 $  and $\tau^2_R $ as defined in \eqref{limit_var2} and (\ref{limit_var2R}), respectively, it holds
\begin{enumerate}
\item[(i)] $ {\rm Var}^*(L_{n}^*) \stackrel{P}{\longrightarrow} \tau^2,$
\item[(ii)] $ \sup_{x \in \R} | P^*(L_n^* \leq x) - P(L_n \leq x) | =o_P(1) $,
\item[(iii)] $ \sup_{x \in \R} | P^*(L_{n,R}^* \leq x) - P(L_{n,R} \leq x) | =o_P(1) $.
\end{enumerate}
\end{theorem}

\

Notice that the above theorems allow for the use of the bootstrap distributions
in order to construct an asymptotic $100(1-2\alpha)$ level confidence interval for $ M(\varphi,f)$ (respectively $ R(\varphi,f)$). To elaborate,  such a confidence interval for $M(\varphi,f)$   is given by
\[ \big(M(\varphi,I_n) - t^*_{1-\alpha}  \frac{1}{\sqrt{n}}\widetilde{s}_n,   \ M(\varphi,I_n) - t^*_{\alpha}\frac{1}{\sqrt{n}}\widetilde{s}_n\big),\]
where  $\widetilde{s}_n^2=Var^*(L_n^*)$ equals
\begin{align*}
\widetilde{s}_n^2 & =\frac{4\pi^2}{b}\sum_{j_1\in  \mathcal{G}(b)}\sum_{j_2\in  \mathcal{G}(b)}\varphi(\lambda_{j_1,b}) \varphi(\lambda_{j_2,b})\widehat{f}_n(\lambda_{j_1,b})\widehat{f}_n(\lambda_{j_2,b})\\
& \ \ \ \ \ \times \frac{1}{N}\sum_{t=1}^N(U_{t,b}(\lambda_{j_1,b})-1)(U_{t,b}(\lambda_{j_2,b})-1),
\end{align*}
and $ t^*_\alpha $ and $ t^*_{1-\alpha}$ are the lower and upper $\alpha$ and $1-\alpha$ quantiles of the distribution of  the studentized quantity $ L^\ast_n/\widetilde{s}_n$.  Similarly,  for $R(\varphi,f)$ the desired confidence interval is given by
 \[ \big(R(\varphi,I_n) - t^*_{R,1-\alpha}  \frac{1}{\sqrt{n}}\widetilde{s}_{n,R},   \ R(\varphi,I_n) - t^*_{R,\alpha}\frac{1}{\sqrt{n}}\widetilde{s}_{n,R}\big),\]
where
\begin{align*}
\widetilde{s}_{n,R}^2 & =\frac{4\pi^2}{b D_b^4}\sum_{j_1\in  \mathcal{G}(b)}\sum_{j_2\in  \mathcal{G}(b)}\widehat{w}(\lambda_{j_1,b}) \widehat{w}(\lambda_{j_2,b})\widehat{f}_n(\lambda_{j_1,b})\widehat{f}_n(\lambda_{j_2,b})\\
& \ \ \ \ \ \times \frac{1}{N}\sum_{t=1}^N(U_{t,b}(\lambda_{j_1,b})-1)(U_{t,b}(\lambda_{j_2,b})-1),
\end{align*}
$D_b=2\pi b^{-1}\sum_{j\in \mathcal{G}(b)} \widehat{f}_n(\lambda_{j,b})$
and $ t^*_{R,\alpha} $ and $ t^*_{R, 1-\alpha}$ are the lower and upper $\alpha$ and $1-\alpha$ quantiles of the distribution of  the studentized quantity $ L^\ast_{n,R}/\widetilde{s}_{n,R}$.

\
Our last theorem establishes consistency of the hybrid bootstrap approaches $ \widetilde{V}_{n}^*$  and $ \widetilde{V}_{n,R}^*$ .
\

\begin{theorem} \label{theorem2}
Let the assumptions of Proposition~\ref{prop1} as well as assertion \eqref{CLT} be fulfilled. Then, with $ \tau^2 $ and $ \tau^2_R$ as in Theorem~\ref{theorem1},
 it holds
\begin{enumerate}
\item[(i)] $ {\rm Var}^*(\widetilde{V}_{n}^*) \stackrel{P}{\longrightarrow} \tau^2,$
\item[(ii)] $ \sup_{x \in \R} | P^*(\widetilde{V}_{n}^* \leq x) - P(L_n \leq x) | =o_P(1) $,
\item[(iii)] $ \sup_{x \in \R} | P^*(\widetilde{V}_{n,R}^* \leq x) - P(L_{n,R} \leq x) | =o_P(1) $.
\end{enumerate}
\end{theorem}
\

\section{Practical Issues and Numerical Results} \label{sec_numresults}
\subsection{Selection of bootstrap parameters} \label{sec_bootpar}
Implementation of  the bootstrap procedure proposed, requires the choice of
 $\widehat{f}_n$ and of the subsampling parameter  $b$. Note that $\widehat{f}_n$ is  an estimator
 of the spectral density $f$,  and that the choice of  the subsampling size   $b$ concerns the estimation of the second variance term
 $ \tau_2$ in (\ref{limit_var2}). In this section    we propose some practical rules for selecting these bootstrap  quantities taking
  into account the aforementioned  different goals.
\

To make the discussion more precise, suppose $\widehat{f}_n(\lambda)$ is a kernel type estimator given by $ \widehat{f}_{n}(\lambda)=(nh)^{-1}\sum_{j\in {\mathcal G}(n)}K_h(\lambda-\lambda_{j,n}) I_n(\lambda_{j,n})$, where $K$ is the smoothing kernel,   $h=h(n)>0$ is a   smoothing bandwidth
and $K_h(\cdot)=h^{-1}K(\cdot/h)$.  Following,  Beltr\~ao and Bloomfield (1987), Hurvich (1985) and Robinson (1991), $h$ can be chosen by minimizing  the cross validation type criterion
 \[ CV(h) = N^{-1}\sum_{j=1}^N \big\{\log (\widehat{f}_{-j}(\lambda_{j,n}) + I_n(\lambda_{j,n})/\widehat{f}_{-j}(\lambda_{j,n})\big\},\]
 over a grid of values of $h$. Here,  $ \widehat{f}_{-j}(\lambda_{j,n}) =(nh)^{-1}\sum_{s\in {\mathcal G}_j(n)}K((\lambda_{j,n}-\lambda_{s,n}/h))I_n(\lambda_{s,n})$ and $ {\mathcal G}_j(n)={\mathcal G}(n)\setminus\{\pm j\}$, that is $ \widehat{f}_{-j}(\lambda_{j,n})$ is the leave one out estimator of the spectral density $f(\lambda_{j,n})$.
 Notice that the same procedure can be used to select the truncation lag $M_n$ in a lag window type estimator of the spectral density $f$; see Robinson (1991).

\

The choice of the block length $b$ is more involved. Recall that the estimator of $\tau_2$ obtained  using the  convolved pseudo
periodograms of subsamples, is given by
\[\widehat{\tau}_{2}(b)=\frac{4\pi^2}{b} \sum_{j_1 \in \mathcal{G}(b)} \sum_{j_2 \in \mathcal{G}(b)\setminus \{j_1,-j_1\}} \varphi(\lambda_{j_1,b}) \, \varphi(\lambda_{j_2,b}) \, \textrm{Cov}^*(I_b^{(1)}(\lambda_{j_1,b}),I_b^{(1)}(\lambda_{j_2,b})), \]
and
\[ \textrm{Cov}^*(I_b^{(1)}(\lambda_{j_1,b}),I_b^{(1)}(\lambda_{j_2,b}))=\widehat{f}_n(\lambda_{j_1,b})
\widehat{f}_n(\lambda_{j_2,b})\frac{1}{N}\sum_{t=1}^N (U_{t,b}(\lambda_{j_1,b})- 1)(U_{t,b}(\lambda_{j_2,b})- 1).\]
One approach could be to choose $b$ by minimizing   the mean square error  $ E(\widehat{\tau}_{2}(b)-\tau_2)^2$.  This   approach is, however,   invisible  since $\tau_2$ is unknown. We, therefore,  propose in the following a simple procedure to determine  $b$ which is
 based  on  the idea to select $b$ in a way which minimizes the aforementioned  mean square error  under the working assumption that  the dependence structure of the underlying process can be  approximated by an autoregressive (AR) model. The following bootstrap based procedure, makes  this idea precise.
\begin{enumerate}
\item Fit to the observed time series $X_1, X_2, \ldots, X_n$ an AR(p) process using Yule-Walker estimators and select  the  order $p$  by minimizing the AIC criterion,  let $\widehat{a}_{j,p}$, $j=1,2, \ldots, p$, be the estimated parameters and denote by $ \widehat{e}_t=\widetilde{e}_t-\overline{e}_n$, $t=p+1, p+2, \ldots, n$, the estimated and centered residuals, i.e., $\widetilde{e}_t= X_t-\sum_{j=1}^p\widehat{a}_{j,p}X_{t-j}$ and $ \overline{e}_n=(n-p)^{-1}\sum_{t=p+1}^n\widetilde{e}_{t}$.
\item Let
\[ \widehat{\tau}_{2,AR}=(\widehat{\eta}_{AR}-3)\Big(\int_{-\pi}^{\pi}\varphi(\lambda) \widehat{f}_{AR}(\lambda)d\lambda\Big)^2,\]
where
\[ \widehat{f}_{AR}(\lambda)=\frac{\widehat{\sigma}^2_e}{2\pi}\big|1-\sum_{j=1}^p \widehat{a}_{j,p}e^{-i \lambda j}\big|^{-2}, \ \ \widehat{\eta}_{AR}= (n-p)^{-1}\sum_{t=p+1}^n\widehat{e}_t^4 \big/ \sigma^4_e, \]
and  $\widehat{\sigma}^2_e =(n-p)^{-1}\sum_{t=p+1}^n\widehat{e}_t^2$.
\item  Generate $\widetilde{X}_1, \widetilde{X}_2, \ldots, \widetilde{X}_n$ using the fitted AR model, that is, $ \widetilde{X}_t=\sum_{j=1}^p\widehat{a}_{j,p} \widetilde{X}_{t-j} + e^+_t$, where
the $ e^+_t$ are i.i.d. having distribution the empirical distribution function of the  centered residuals $ \widehat{e}_t$, $t=p+1, \ldots, n$.
\item Calculate  for each $ b \in \{b_{\min}, b_{\min+1}, \ldots, b_{\max}\}$ the estimator
\[  \widehat{\tau}_{2,AR}(b)=\frac{4\pi^2}{b} \sum_{j_1 \in \mathcal{G}(b)} \sum_{j_2 \in \mathcal{G}(b)\setminus \{j_1,-j_1\}} \varphi(\lambda_{j_1,b}) \, \varphi(\lambda_{j_2,b}) \, \textrm{Cov}^*(\widetilde{I}_b^{(1)}(\lambda_{j_1,b}),\widetilde{I}_b^{(1)}(\lambda_{j_2,b})),\]
where $ \widetilde{I}_b^{(1)}(\lambda) =(2\pi b)^{-1} |\sum_{s=1}^{b} \widetilde{X}_{i_1^*+s-1} \, e^{-i\lambda_{j,b} s}|^2$ and $ i_1^*$ is a random variable
having the discrete uniform distribution on the set $\{1,2, \ldots, N\}$.
\item Repeat (3) and (4) a large number of, say,  $ L$ times and calculate for every $ b\in \{b_{\min}, b_{\min+1}, \ldots, b_{\max}\}$ the empirical mean square error $ V_L(b)=L^{-1}\sum_{l=1}^L\big(\widehat{\tau}^{(l)}_{2,AR}(b)- \widehat{\tau}_{2,AR}\big)^2 $, where $ \widehat{\tau}^{(l)}_{2,AR}(b)$ is the value of $  \widehat{\tau}_{2,AR}(b)$ obtained in the $l$th repetition.  Select $b^*$ as $ b^*=\min_{\{b_{\min}, b_{\min+1}, \ldots, b_{\max}\}}V_L(b)$.
\end{enumerate}

\subsection{Simulation results} \label{sec_sim}

We investigate the finite sample performance of the new approach based on convolved bootstraped periodograms (CBP) and we compare it with that of the hybrid periodogram bootstrap procedure (HPB) and of the multiplicative periodogram bootstrap (MPB). For this we consider time series of  different lengths stemming from the following four time series models with parameters $ \phi=\theta=-0.7$:
\begin{enumerate}
\item[] {\bf Model I:} \ $X_t = \theta \varepsilon_{t-1} + \varepsilon_t$, \  and  i.i.d. innovations $ \varepsilon_t\sim Exp(1)-1$.
\item[] {\bf Model II:}  \ $ X_t = v_t + \theta v_{t-1} $,  \ $ v_t = \big(\varepsilon_t\sqrt{a_0+a_1 v^2_{t-1}}\big)/\sqrt{0.6}$, \ $a_0=0.3$, $a_1=0.5$  and  i.i.d. $ \varepsilon_t\sim \mathcal N(0,1)$.
\item[] {\bf Model III:}\  $ X_t = z_t + \theta z_{t-1}$, \   $ z_{t}=\varepsilon_{t}\cdot \varepsilon_{t-1}$ and  $ \varepsilon_t $  as for Model II.
\item[] {\bf Model IV:}  \ $ X_t = \phi \sin(X_{t-1}) + \varepsilon_t $ and $ \varepsilon_t$ as for Model II.
\end{enumerate}
The above choices  cover a wide range of linear and nonlinear models commonly used in time series analysis. In particular, Model I is a linear MA(1) with centered, exponentially distributed i.i.d. innovations, Model II is a MA(1) model  with ARCH(1) innovations,  Model III is a bilinear model and Model IV is a nonlinear AR model.

\

We first demonstrate the capability  of the new bootstrap procedure  to  consistently estimate the distribution of  interest and investigate its sensitivity with respect to the choice of the  size $b$ of the subsamples used. For this  we consider
the distribution of the first order sample autocovariance $ \sqrt{n}(\widehat{\gamma}(1)-\gamma(1))$. Time series of lengths  $n=150$ and $n=2000$ are considered,  
where the later sample size  has been chosen in order to clearly see 
 the differences in the consistency behavior of  the different bootstrap  procedures. 
 In order to measure  the closeness of the bootstrap distribution  to the exact distribution of interest, we calculated    the $d_1$ distance between distributions defined  as   $d_1(F,G)=\int_{0}^{1}|F^{-1}(x)-G^{-1}(x)|dx$, for $F$ and $G$ distribution functions. Figure 1 shows  mean  distances calculated over $R=200$ repetitions using  a wide range of different  subsampling  sizes $b$. To evaluate  the exact distribution of $ \sqrt{n}(\widehat{\gamma}(1)-\gamma(1))$, 10,00 replications have been used while all  bootstrap approximations are based on  1,000 replications.
To estimate the spectral density $f$ used in the multiplicative as well as in the new periodogram bootstrap procedures, the  Parzen lag window, see Priestley (1981),
 has been used
 with  truncation lag $ M_n=15$  for $n=150$ and $ M_n=25$ for $n=2000$.  
 
 \
 
The results obtained are shown in  Figure 1 and Figure 2. 
As it can be seen, the differences between the method generating independent periodogram ordinates and those capturing the dependence of the periodogram are not large  for the sample size of $n=150$ observations with the estimation results being rather mixed and those of the moving average model with 
ARCH innovations showing a larger sensitivity with respect to the value of the subsampling parameter $b$.  However, the differences between the different methods are clearly  seen  for the sample size  of $n=2000$ observations. Here  the  distances of the MPB estimates are more or less the same as those for $n=150$, while due to their consistency, the better  performance  
of  the CBP and the HPB  methods  in approximating the distributions of interest are  clearly  highlighted in the corresponding exhibits.
In these exhibits and apart from small values of $b$, the   good behavior of these methods   seems
to be valid   for a wide range of  values of the  subsampling parameter $b$.  Furthermore, the hybrid periodogram bootstrap delivers good estimation results   even for small values of $b$ making this bootstrap procedure less sensitive with respect to the choice of this  parameter.  Notice that  the  bootstrap estimates  for both aforementioned  bootstrap methods  are,  for sufficiently large values of $b$, very similar  and closely follow each other for the different values of $b$ considered.

\begin{figure}[ht] \label{fig.dens}
\begin{center}
\includegraphics[angle=0,height=7cm,width=7.0cm]{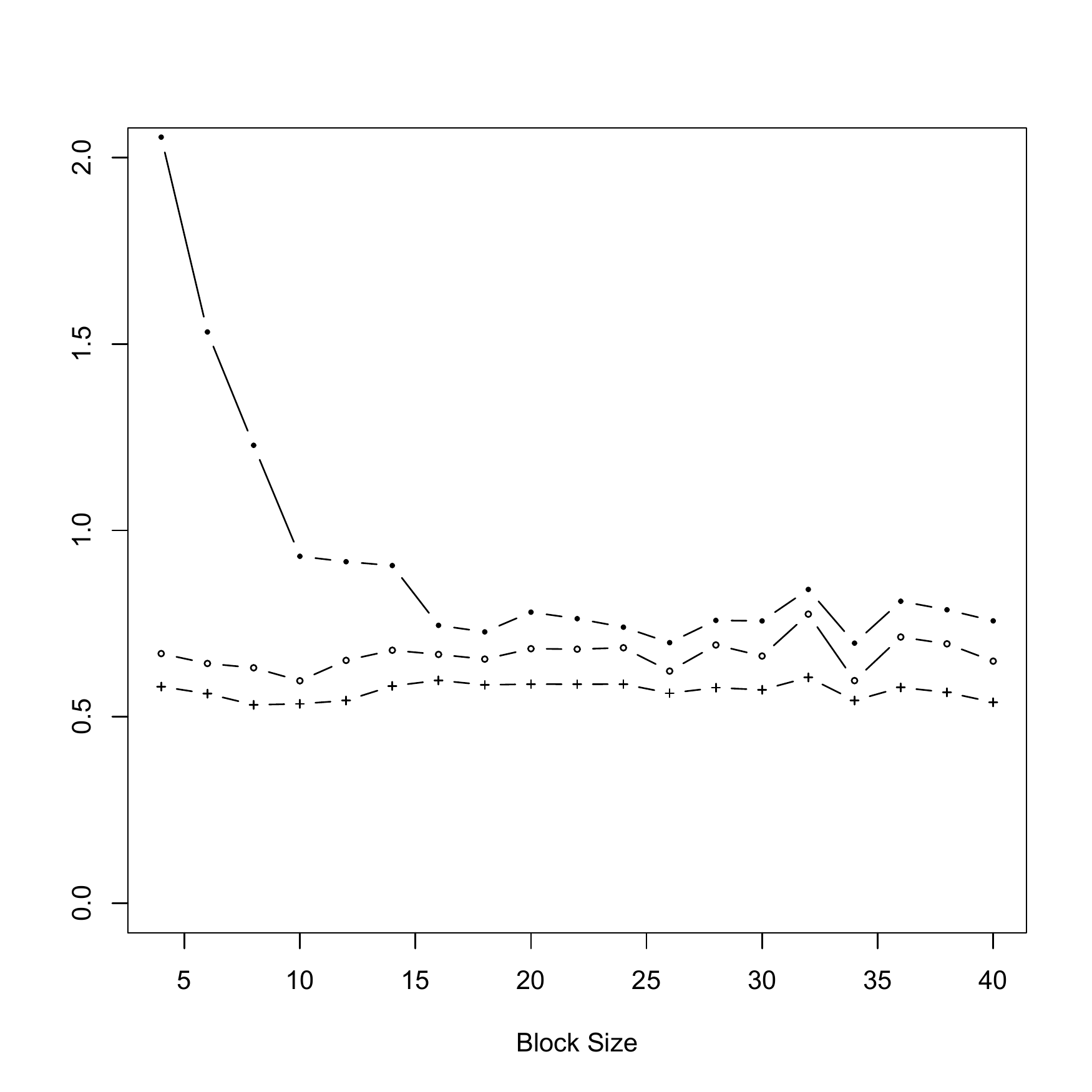}
\includegraphics[angle=0,height=7cm,width=7.0cm]{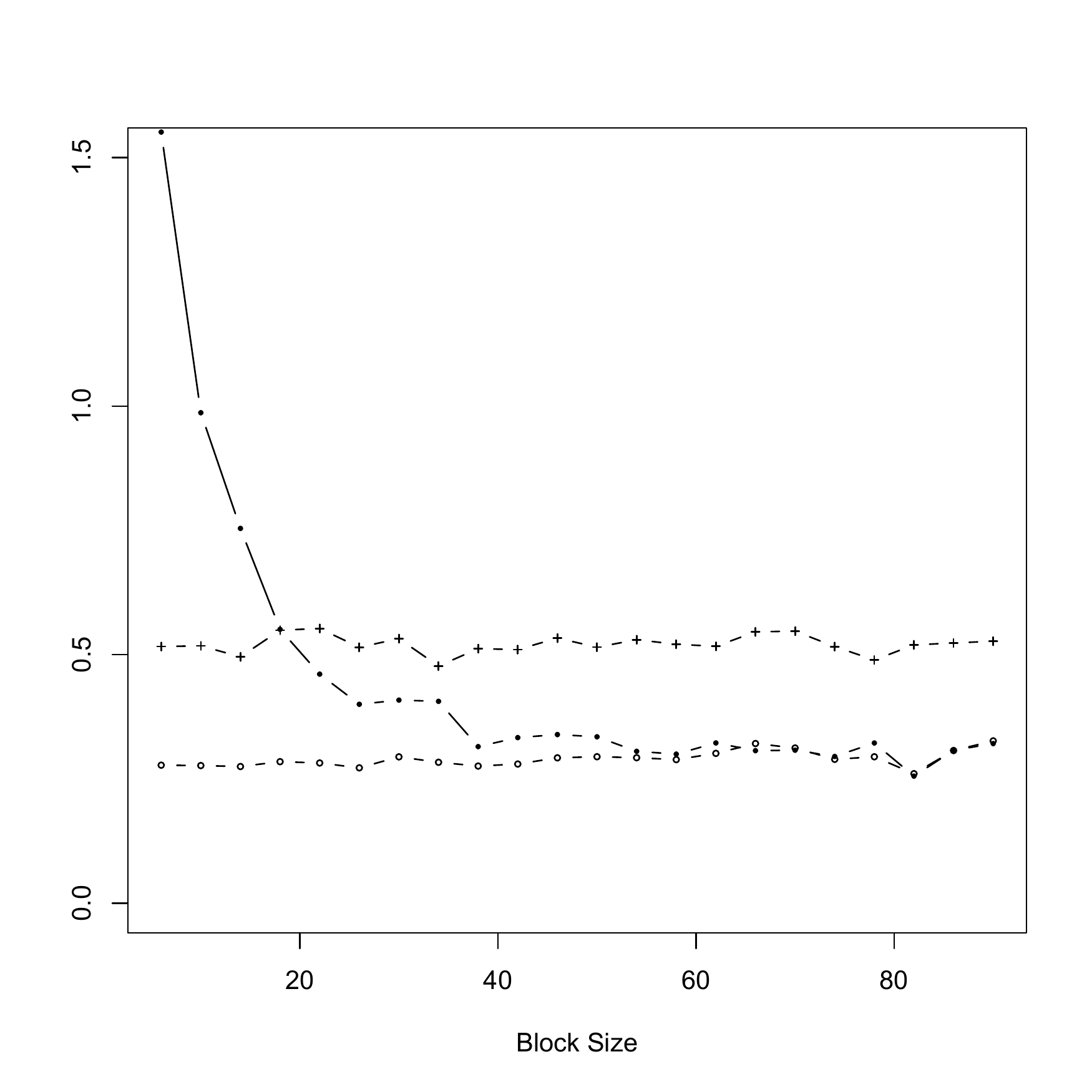}\\
\includegraphics[angle=0,height=7cm,width=7.0cm]{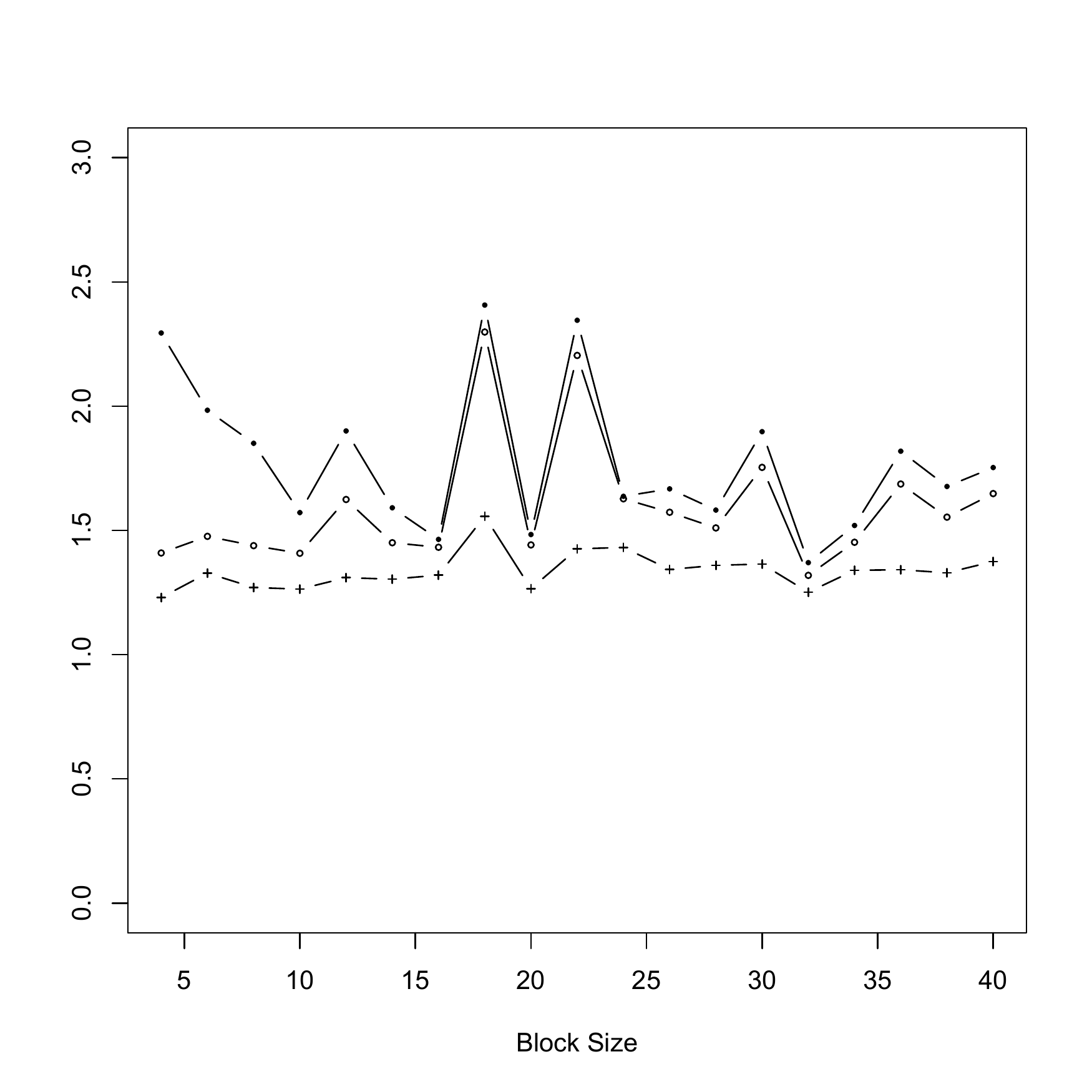}
\includegraphics[angle=0,height=7cm,width=7.0cm]{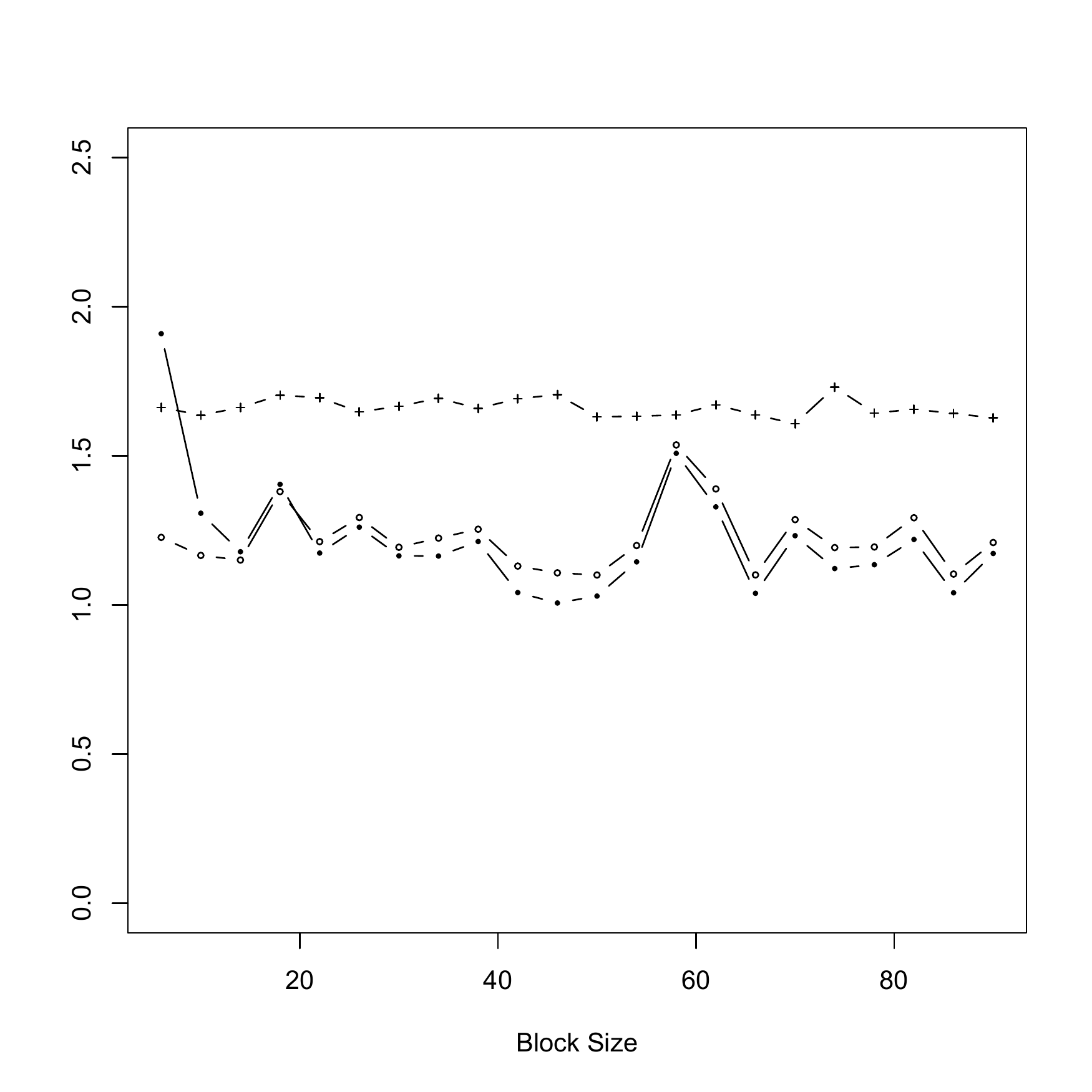}
\end{center}
\caption{ Average $d_1$-distances  between the exact and the bootstrap distribution of $ \sqrt{n}(\widehat{\gamma}(1)-\gamma(1))$ for various block sizes and for Model I (first row) and 
Model II (second raw).  The left panels refer to  $n=150$ and the right panels to $n=2000$.   The crosses denote the  $d_1$-distance of  the multiplicative periodogram bootstrap,  the dots of the convolved bootstraped periodograms and the  circles of  the hybrid periodogram bootstrap.}
\end{figure}

\begin{figure}[ht] \label{fig.dens}
\begin{center}
\includegraphics[angle=0,height=7cm,width=7.0cm]{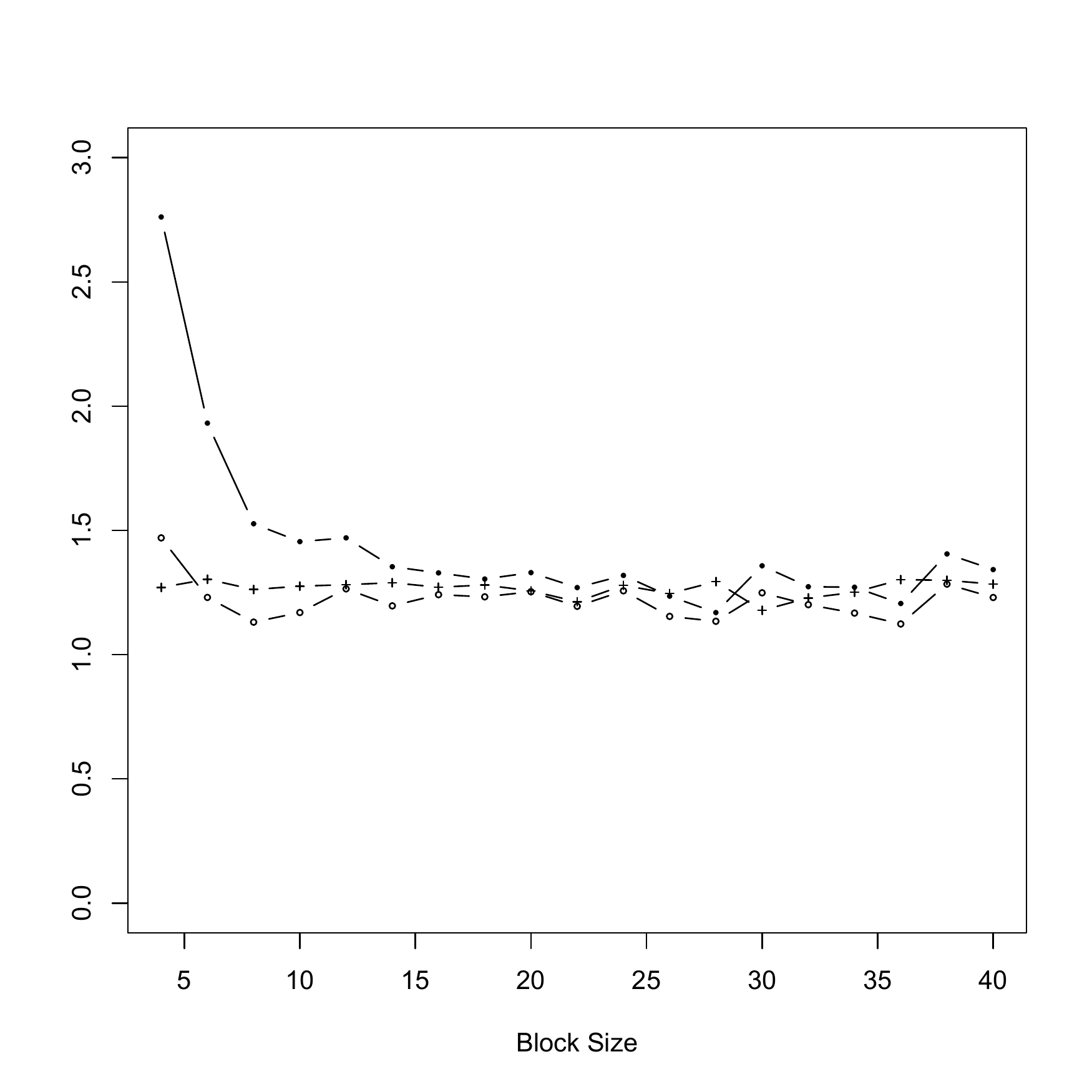}
\includegraphics[angle=0,height=7cm,width=7.0cm]{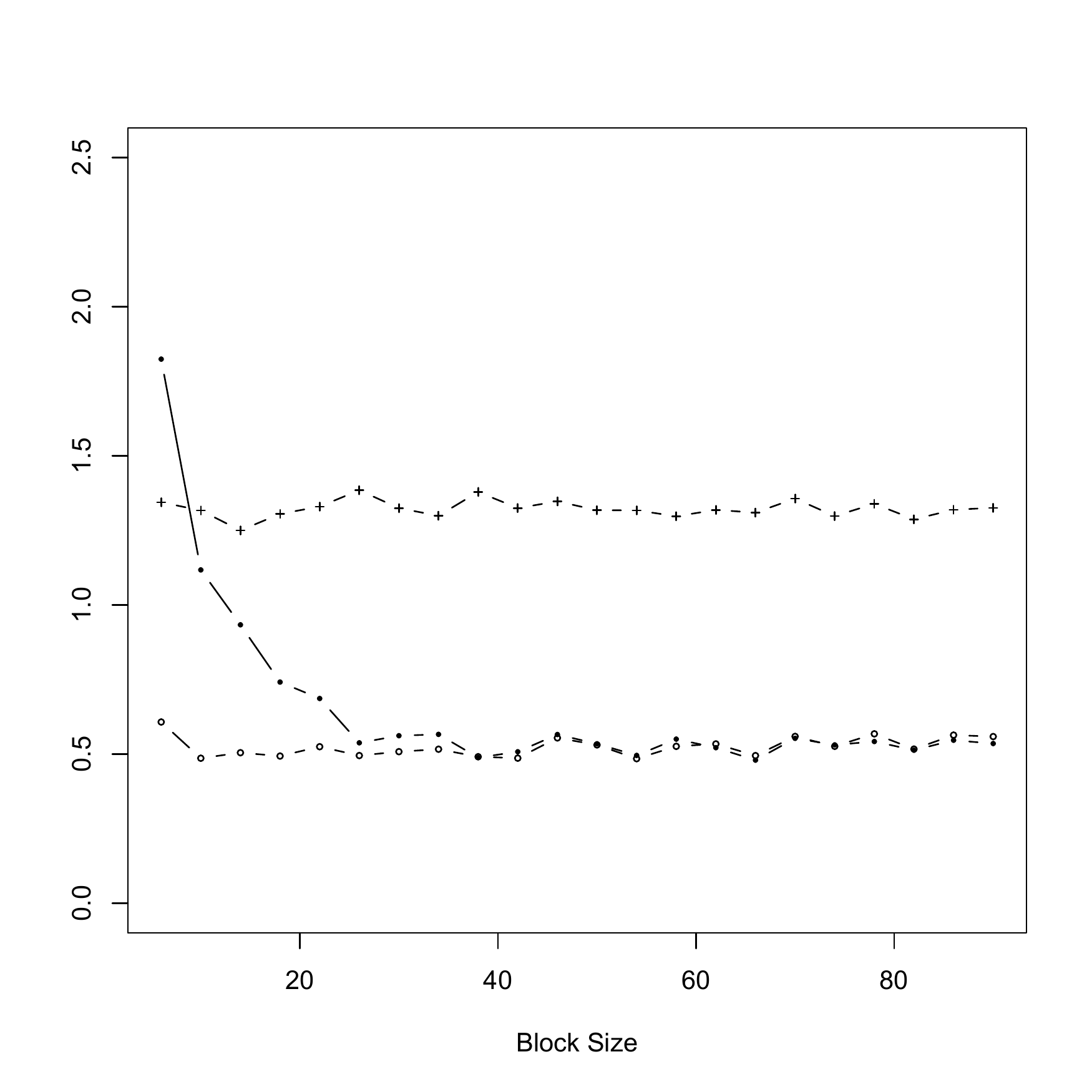}\\
\includegraphics[angle=0,height=7cm,width=7.0cm]{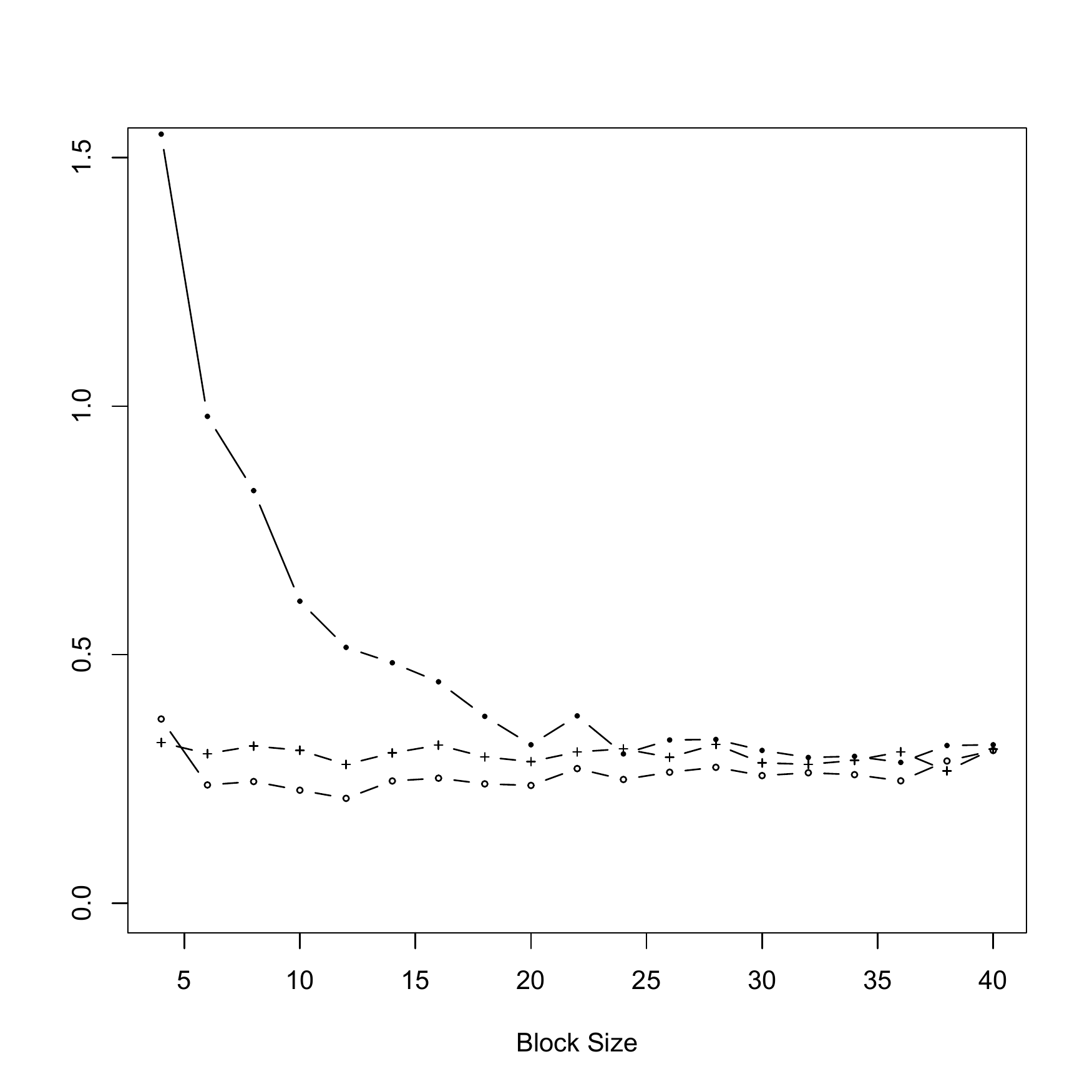}
\includegraphics[angle=0,height=7cm,width=7.0cm]{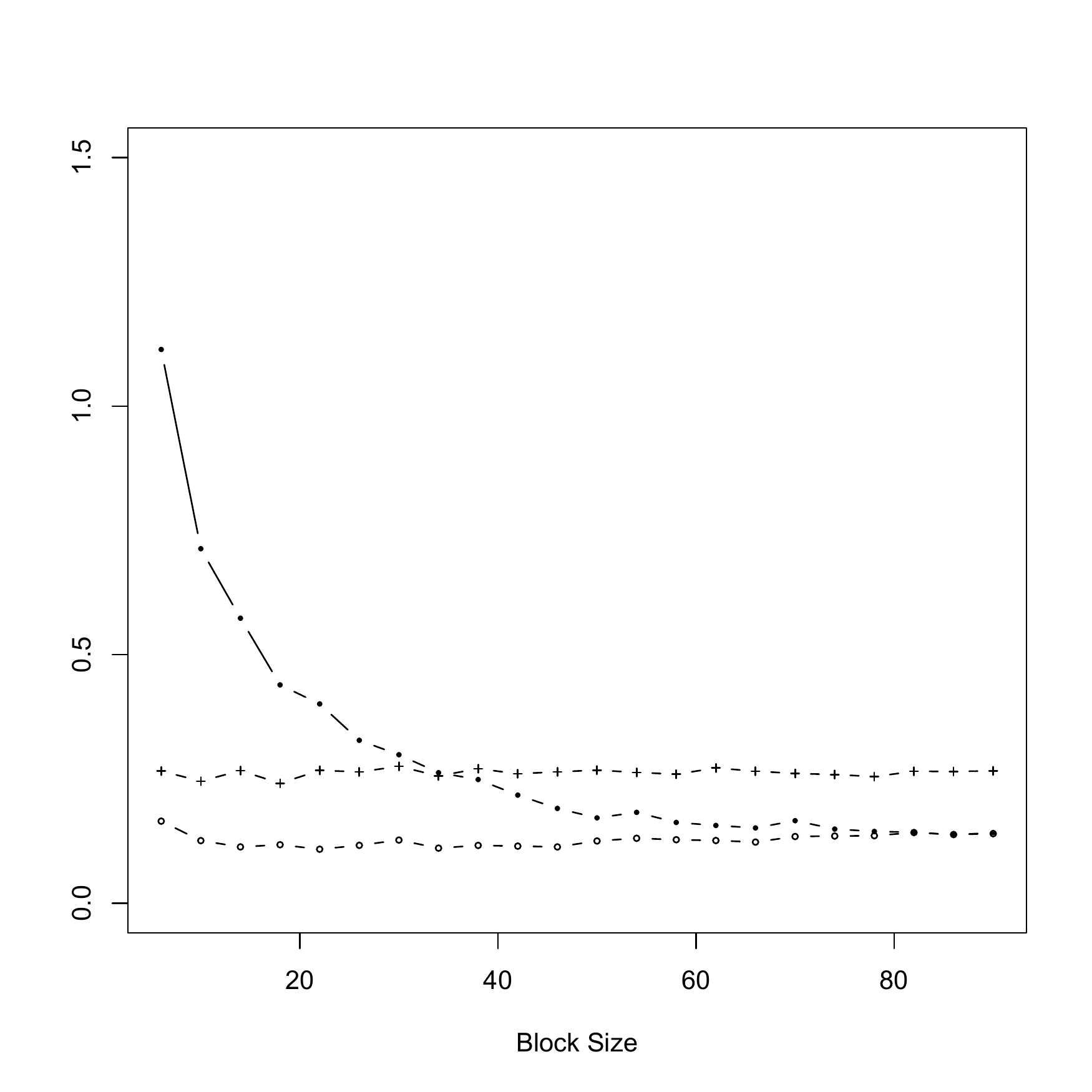}
\end{center}
\caption{Average $d_1$-distances  between the exact and the bootstrap distribution of $ \sqrt{n}(\widehat{\gamma}(1)-\gamma(1))$ for various block sizes and 
Model III (first raw) and Model IV (second raw).  The left panels refer to  $n=150$ and the right panels to $n=2000$.   The crosses denote the  $d_1$-distance of  the multiplicative periodogram bootstrap,  the dots of the convolved bootstraped periodograms and the  circles of  the hybrid periodogram bootstrap.}
\end{figure}

\

We next  investigate the finite sample performance of the convolved periodogram bootstrap to estimate  the standard deviation of the first order sample 
autocorrelation $\sqrt{n}\widehat{\rho}(1)$. 
We report the estimated exact standard deviation of $\sqrt{n}\widehat{\rho}(1)$ calculated over 50,000 repetitions, the mean and the standard deviation of the bootstrap estimates calculated over 500 replications based on $M=500$ bootstrap samples and for a range of different values of the subsampling parameter $b$. The value of the truncation lag used to obtain the spectral density estimator $\widehat{f}_n$ using the Parzen lag window  has been set equal to $15, 20$ and $25$ for the sample sizes $n=150$, $n=300$ and $n=500$ respectively. The numerical results obtained are reported in Table 1. As it is seen from this table, the bootstrap estimates are good and  they improve as the sample size increases. They  also seem to be not very  sensitive with respect to the choice of the subsampling parameter $b$. Notice that this  behavior 
has been also observed  by  the   results presented in Figure 1 and 2.

\clearpage

\begin{scriptsize}
\begin{table}[ht]
\begin{tabular}{c|cccc|cccc|cccc}
\hline
  & &  &  &  &  &  &      &  &  &  & &  \\
   & n=150 &  &  &  &n=300 &  &  &  &n=500 &  &  \\
   & Est.Ex. & b & Mean & Std & Est.Ex. & b & Mean & Std  & Est.Ex. & b & Mean & Std\\
 \hline
  M I    &0.7302            &18  &0.7418 & 0.1725    &  0.7316 & 20 & 0.7465 & 0.1607 &  0.7291 & 20 & 0.7320 & 0.1239 \\
           &                       &20  & 0.7401 & 0.1904     &            & 22 & 0.7560 & 0.1575 &                & 25 & 0.7192 & 0.1213 \\
           &                       &22  & 0.7178 & 0.1721     &            & 24 & 0.7433 & 0.1624 &               & 30 & 0.7302 & 0.1449\\
     &  &  &  &  &  &      &  &  &  & & &  \\
  M II    &1.0201            &18  &0.9004 & 0.2465     & 1.0946 & 20 & 0.9670 & 0.2852 &  1.1493 & 20 & 1.0254 & 0.2500 \\
            &                       &20  & 0.9164 & 0.2151    &             & 22 & 0.9923 & 0.2692 &              & 25 & 0.9841 & 0.2654\\
            &                       &22  & 0.9252 & 0.2715    &             & 24 & 0.9817 & 0.2769&               & 30 & 1.0115 & 0.2685\\
        &  &  &  &  &  &    &  &  &  &  & &  \\
  M III  &1.0187            &18   & 0.9436 & 0.2641     & 1.0427&  20 & 0.9643 & 0.2260 &  1.0527 &20  & 1.0391 & 0.2419 \\
           &                       &20   & 0.9792 & 0.2563     &            & 22 & 1.0175 & 0.2398  &             &25&  0.9794 & 0.2064\\
            &                       &22  & 0.9405 & 0.2777     &            & 24 & 0.9785 & 0.2261  &             &30 & 1.0041 & 0.2169\\
              &  &  &  &  &  &      &  &  &  & &  \\
  M IV    &0.8395            &18  &0.8387 & 0.1761     & 0.8435& 20 & 0.8524 & 0.1386 &  0.8439 & 20 & 0.8531 & 0.1128 \\
             &                       &20  & 0.8361 & 0.1821    &           & 22 & 0.8602 & 0.1529  &            &25    &  0.8258 & 0.1099\\
             &                       &22  & 0.8583 & 0.2008    &           & 24 & 0.8594 & 0.1518 &             & 30& 0.8514 & 0.1393\\
\hline
\end{tabular}
\label{table1}
\vspace*{0.3cm}
\caption{Bootstrap estimates of the standard deviation of  the first order autocorrelation estimator $ \sqrt{Var(\sqrt{n}\widehat{\rho}(1))}$.  ``Est.Ex.'' refers to the estimated exact standard deviation, ``Mean''  to the average  and ``Std'' to the standard deviation of the bootstrap estimates.}
\end{table}
\end{scriptsize}
%
%

\section*{Appendix}

{\sc Proof of Lemma \ref{lemma1} $ (i) $:} As a preliminary, a straightforward calculation yields for any Fourier frequency $ \lambda_{j,b} $
\be
& & \big|{\rm Cov}(I_{t_1,b}(\lambda_{j,b}), I_{t_2,b}(\lambda_{j,b})) \big| \nonumber \\
&\leq & \frac{1}{4\pi^2 b^2} \sum_{l_1,l_2,l_3,l_4=1}^b |{\rm Cov}(X_{t_1+l_1-1}X_{t_1+l_2-1}, X_{t_2+l_3-1}X_{t_2+l_4-1}) |\,. \label{MMlem41}
\ee
Using this, we have
\begin{align*}
{\rm Var}(\widetilde{f}(\lambda_{j,b})) & = \frac{1}{N^2}\sum_{t_1=1}^N\sum_{t_2=1}^N {\rm Cov}(I_{t_1,b}(\lambda_{j,b}), I_{t_2,b}(\lambda_{j,b}))\\
& \leq  \frac{1}{4\pi^2 b^2 N^2}\sum_{t_1,t_2=1}^N\sum_{l_1,l_2,l_3,l_4=1}^b\\
&\ \  \Big\{ \big|\gamma((t_2-t_1)+(l_3-l_1))\big|\big|\gamma((t_2-t_1)+(l_4-l_2))\big|\\
& \ \ + \big|\gamma((t_2-t_1)+(l_4-l_1))\big|\big|\gamma((t_2-t_1)+(l_3-l_2))\big|\\
& \ \ + \big|{\rm cum}(X_{t_1+l_1-1},X_{t_1+l_2-1},X_{t_2+l_3-1}, X_{t_2+l_4-1})\big|\Big\}\\
& = G_{1,n}+G_{2,n}+G_{3,n}
\end{align*}
with an obvious notation for $G_{j,n}$, $j=1,2,3$, and where all three terms are uniform over all frequencies $ (\lambda_{j,b})_{j \in \mathcal{G}(b)} $.  Now, because $ \sum_{h\in\Z}|\gamma(h)| <\infty$ we get by summing first over $l_4$ and then over $t_1$, that $ | G_{1,n}|  = \mathcal{O}(b/N)$. For the second term summing first over $l_3$ and then over $t_1$ we also get $ |G_{2,n}|=\mathcal{O}(b/N)$. Finally, by the summability of the fourth order cumulants we get for the last term after summing first over $l_2$ then over $l_3$ and then over $t_1$, that $ |G_{3,n}| =\mathcal{O}(1/N)$.
With $ E(\widetilde{f}(\lambda_{j,b}))=E(I_{1,b}(\lambda_{j,b})) $ and $ |\mathcal{G}(b)|\leq b $, we have
\bee
E \Big( \sum_{j \in \mathcal{G}(b)} \big| \widetilde{f}(\lambda_{j,b}) - E I_{1,b}(\lambda_{j,b}) \big| \Big) \leq  \sum_{j \in \mathcal{G}(b)}  \big({\rm Var}(\widetilde{f}(\lambda_{j,b}))\big)^{1/2} \leq b \, \mathcal{O}\big( \sqrt{b/N} \big)\,,
\eee
and assertion $(i)$ follows. $ \hfill \square $

\

{\sc Proof of Lemma \ref{lemma1} $ (ii) $:} We use the following  decomposition:
\begin{align*}
\frac{1}{N}\sum_{t=1}^N &\big(I_{t,b}(\lambda_{j,b})I_{t,b}(\lambda_{s,b}) - E( I_{t,b}(\lambda_{j,b})I_{t,b}(\lambda_{s,b})) \big) \\
&  =\frac{1}{N}\sum_{t=1}^NE(I_{t,b}(\lambda_{s,b}))\big(I_{t,b}(\lambda_{j,b})- E (I_{t,b}(\lambda_{j,b})) \big) \\
& \ \ + \frac{1}{N}\sum_{t=1}^NE(I_{t,b}(\lambda_{j,b}))\big(I_{t,b}(\lambda_{s,b})- E (I_{t,b}(\lambda_{s,b})) \big) \\
& \ \ +  \frac{1}{N}\sum_{t=1}^N\Big\{\big(I_{t,b}(\lambda_{j,b})- E (I_{t,b}(\lambda_{j,b})) \big)\big(I_{t,b}(\lambda_{s,b})- E (I_{t,b}(\lambda_{s,b})) \big)\\
& \ \ \ \ \ \ -E\Big[\big(I_{t,b}(\lambda_{j,b})- E (I_{t,b}(\lambda_{j,b})) \big)\big(I_{t,b}(\lambda_{s,b})- E (I_{t,b}(\lambda_{s,b})) \big)\Big]\Big\}\\
& = K_{j,s,n}^{(1)}+K_{j,s,n}^{(2)}+K_{j,s,n}^{(3)},
\end{align*}
with an obvious notation for $ K_{j,s,n}^{(l)}$, $l=1,2,3$. Since
\[ E(I_{t,b}(\lambda_{j,b})) -f(\lambda_{j,b}) = -\frac{1}{2\pi}\sum_{|h|<b}\frac{h}{b} \gamma(h)e^{-i\lambda_{j,b} h} -\frac{1}{2\pi}\sum_{|h|\geq b}\gamma(h)e^{-i\lambda_{j,b}h}=:C_b(\lambda_{j,b})\,, \]
we get for $K_{j,s,n}^{(1)}$ that
\begin{align*}
\sum_{j,s \in \mathcal{G}(b)} |K_{j,s,n}^{(1)}| & = \sum_{s \in \mathcal{G}(b)} \big|f(\lambda_{s,b}) +C_b(\lambda_{s,b})\big| \, \sum_{j \in \mathcal{G}(b)} \Big|\frac{1}{N}\sum_{t=1}^N\big(I_{t,b}(\lambda_{j,b})- E (I_{t,b}(\lambda_{j,b})) \Big| \\
&=\mathcal{O}_P\big(\sqrt{b^5/N}\big),
\end{align*}
because of assertion $(i)$ and the fact that $ |C_b(\lambda_{j,b})|=\mathcal{O}(b^{-1}) \rightarrow 0$ as $ b \rightarrow \infty$, uniformly in $\lambda_{j,b}$. Notice that $\sum_{j,s} |K_{j,s,n}^{(2)}| =\mathcal{O}_P\big(\sqrt{b^5/N}\big)$ by the same arguments. For the centered expression $ K_{j,s,n}^{(3)} $ we have
\bee
E \left(\sum_{j,s \in \mathcal{G}(b)} |K_{j,s,n}^{(3)}| \right) \leq \sum_{j,s \in \mathcal{G}(b)} \sqrt{{\rm Var}(K_{j,s,n}^{(3)})} \,,
\eee
thus, the proof can be completed by showing $ {\rm Var}(K_{j,s,n}^{(3)})\leq Cb/N $ for some constant $ C<\infty $, not depending on $ j,s $ and $ n $. Using the notation $ I^c_{t,b}(\lambda_{j,b}):= I_{t,b}(\lambda_{j,b})- E (I_{t,b}(\lambda_{j,b}))$ we get
\begin{align*}
{\rm Var}\big(K_{j,s,n}^{(3)}\big) & = \frac{1}{N^2}\sum_{t_1=1}^N\sum_{t_2=1}^N\Big\{ E( I^c_{t_1,b}(\lambda_{j,b}) I^c_{t_2,b}(\lambda_{j,b}) )E( I^c_{t_1,b}(\lambda_{s,b}) I^c_{t_2,b}(\lambda_{s,b}) ) \\
& \ \  \ \ \ \ + E( I^c_{t_1,b}(\lambda_{j,b}) I^c_{t_2,b}(\lambda_{s,b}))E( I^c_{t_1,b}(\lambda_{s,b}) I^c_{t_2,b}(\lambda_{j,b})) \\
  & \ \ \ \ \  \  + {\rm cum}( I^c_{t_1,b}(\lambda_{j,b}),I^c_{t_1,b}(\lambda_{s,b}),I^c_{t_2,b}(\lambda_{j,b}),I^c_{t_2,b}(\lambda_{s,b}))\Big\}\\
  & = S_{j,s,n}^{(1)}+S_{j,s,n}^{(2)}+S_{j,s,n}^{(3)},
\end{align*}
with an obvious notation for $ S_{j,s,n}^{(l)}$, $l=1,2,3$. $ S_{j,s,n}^{(1)}$ and $ S_{j,s,n}^{(2)}$ are handled similarly, so we consider only $S_{j,s,n}^{(1)}$.  Evaluating the expectations involved, and using \eqref{MMlem41}, we get for this term the bound
\begin{align*}
|S_{j,s,n}^{(1)}| & \leq \frac{1}{16\pi^4 b^4 N^2} \sum_{t_1,t_2=1}^N\Big[\sum_{s_1,s_2,s_3,s_4=1}^b\Big(|\gamma((t_2-t_1)+(s_3-s_1))\gamma((t_2-t_1)+(s_4-s_2))| \\
& \ \ \ \ + |\gamma((t_2-t_1)+(s_4-s_1))\gamma((t_2-t_1)+(s_3-s_2))|  \\
& \ \ \ \  +  |{\rm cum}(X_0,X_{s_2-s_1},X_{t_2-t_1+s_3-s_1},X_{t_2-t_1+s_4-s_1})|\Big]^2 \,,
\end{align*}
which does not depend on $ j $ and $ s $. Evaluating the above expression we see that three different types of terms appear, which are of the following form
\begin{align} \label{eq.term1}
\frac{1}{16\pi^4 b^4 N^2}\sum_{t_1,t_2}\sum_{s_1\cdots s_4}& \sum_{r_1\cdots r_4} |\gamma((t_2-t_1)+(s_3-s_1))| |\gamma((t_2-t_1)+(s_4-s_2))| \nonumber \\
& \ \ \ \ \times|\gamma((t_2-t_1)+(r_3-r_1))| | \gamma((t_2-t_1)+(r_4-r_2))|,
\end{align}
\begin{align} \label{eq.term2}
\frac{1}{16\pi^4 b^4 N^2}\sum_{t_1,t_2}\sum_{s_1\cdots s_4}& \sum_{r_1\cdots r_4} |\gamma((t_2-t_1)+(s_3-s_1))| |\gamma((t_2-t_1)+(s_4-s_2))| \nonumber \\
& \ \ \ \ \times |{\rm cum}(X_0,X_{r_2-r_1},X_{t_2-t_1+r_3-r_1},X_{t_2-t_1+r_4-r_1})|\,,
\end{align}
and
\begin{align} \label{eq.term3}
\frac{1}{16\pi^4 b^4 N^2}\sum_{t_1,t_2}\sum_{s_1\cdots s_4}& \sum_{r_1\cdots r_4}  |{\rm cum}(X_0,X_{s_2-s_1},X_{t_2-t_1+s_3-s_1},X_{t_2-t_1+s_4-s_1})| \nonumber \\
& \ \ \ \ \times |{\rm cum}(X_0,X_{r_2-r_1},X_{t_2-t_1+r_3-r_1},X_{t_2-t_1+r_4-r_1})|.
\end{align}
For the term (\ref{eq.term1}) we get summing first with respect to  $r_2$, then with respect to  $r_3$, then with respect to  $s_2$ and finally with respect to $t_1$, that this term is $\mathcal{O}(b/N)$. For (\ref{eq.term2}) summing first with respect to $r_4$, then with respect to $r_3$, then with respect to $r_2$, then with respect to $s_4$, and finally with respect to  $t_1$, we get that  this term is $ \mathcal{O}(1/N)$. Finally for the term (\ref{eq.term3}) we get summing first with respect to $r_4$, then with respect to $r_3$, then with respect to $r_2$, then with respect to $s_4$, then with respect to $s_2$ and finally with respect to $ t_1$, that this term is $\mathcal{O}(1/(Nb))$. From this we conclude that  $ S_{j,s,n}^{(1)}=\mathcal{O}(b/N)$ uniformly in $ j,s $.

For the term $S_{j,s,n}^{(3)}$  we get evaluating the expectations involved the bound
\begin{align*}
&| S_{j,s,n}^{(3)}|  \leq \frac{1}{16 \pi^4 b^4N^2}\sum_{t_1,t_2=1}^N \sum_{r_1,r_2, \ldots, r_8=1}^b \\
& \big|{\rm cum}(X_{t_1+r_1-1}X_{t_1+r_2-1},X_{t_1+r_3-1}X_{t_1+r_4-1},  X_{t_2+r_5-1}X_{t_2+r_6-1}, X_{t_2+r_7-1}X_{t_2+r_8-1})\big|.
\end{align*}
Using the Theorem 2.3.2 of Brillinger (1981), we can write the cumulant term as $ \sum_{\nu}{\rm cum}( X_{t}; t\in \nu_1)\cdots {\rm cum}(X_{t}; t \in \nu_p)$ where
the summation is over all indecomposable partitions $ \nu=\nu_1\cup \nu_2 \cup \cdots  \nu_p $ of the two dimensional table
\[\begin{array}{cc} t_1+r_1-1  & t_1+r_2-1 \\ t_1+r_3-1 &  t_1+r_4-1 \\ t_2+r_5-1 &  t_2+r_6-1 \\ t_2+r_7-1 &  t_2+r_8-1.\end{array} \]
Evaluating these cumulant terms it turns out that the above bound for $ S_{3,n}$ is dominated by the cases where only pairs in the corresponding cumulant terms appear, with a typical one given by
\begin{align*}
\frac{1}{b^4N^2} & \sum_{t_1,t_2=1}^N \sum_{r_1,r_2, \ldots, r_8 =1}^b   |{\rm cum}(X_{t_1+r_1-1},X_{t_2+r_6-1})| | {\rm cum}(X_{t_1+r_3-1},X_{t_2+r_8-1})|\\
& \ \ \ \ \ \ \times  |{\rm cum}(X_{t_2+r_5-1},X_{t_1+r_2-1}) | |{\rm cum}(X_{t_2+r_7-1},X_{t_1+r_4-1})|\\
& =\frac{1}{b^4N^2} \sum_{t_1,t_2=1}^N \sum_{r_1,r_2, \ldots, r_8 =1}^b |\gamma((t_2-t_1)+(r_6-r_1))||\gamma((t_2-t_1)+(r_8-r_3))|\\
& \ \ \ \  \ \  \times |\gamma((t_2-t_1)+(r_5-r_2))||\gamma((t_2-t_1)+(r_7-r_4))|\\
& = \mathcal{O}(b/N),
\end{align*}
where the last equality follows by summing first with respect to $r_3$,  then with respect to $r_5$, then with respect to  $r_7$ and finally
with respect to $ t_1$. $ \hfill \square $

\

{\sc Proof of Lemma \ref{lemma2} $ (i) $:} Strict stationarity of $ (X_t)_{t \in \Z} $ implies $ E(W_{1,b})=0 $. Thus, $ E(W_{1,b}^2)=\textrm{Var}(W_{1,b}) $ and this expression equals
\be
& & \frac{(2\pi)^2}{b} \sum_{j_1,j_2 \in \mathcal{G}(b)} \varphi(\lambda_{j_1,b}) \varphi(\lambda_{j_2,b}) \, \textrm{Cov}\big( I_{1,b}(\lambda_{j_1,b}) - \widetilde{f}_b(\lambda_{j_1,b}), I_{1,b}(\lambda_{j_2,b}) - \widetilde{f}_b(\lambda_{j_2,b}) \big) \nonumber \\
&=& \frac{(2\pi)^2}{b} \sum_{j_1,j_2 \in \mathcal{G}(b)} \varphi(\lambda_{j_1,b}) \varphi(\lambda_{j_2,b}) \, ( A_{j_1,j_2,b} + B_{j_1,j_2,b} +C_{j_1,j_2,b} ) \, , \label{MMlem21}
\ee
where
\bee
A_{j_1,j_2,b} &:=& \textrm{Cov}\big( I_{1,b}(\lambda_{j_1,b}) - f(\lambda_{j_1,b}), I_{1,b}(\lambda_{j_2,b}) - f(\lambda_{j_2,b}) \big) \,,\\
B_{j_1,j_2,b} &:=& \textrm{Cov}\big( f(\lambda_{j_1,b}) - \widetilde{f}_b(\lambda_{j_1,b}), I_{1,b}(\lambda_{j_2,b}) - f(\lambda_{j_2,b}) \big) \,,\\
C_{j_1,j_2,b} &:=& \textrm{Cov}\big( I_{1,b}(\lambda_{j_1,b}) - \widetilde{f}_b(\lambda_{j_1,b}), f(\lambda_{j_2,b}) - \widetilde{f}_b(\lambda_{j_2,b}) \big) \,.
\eee
Now, since $ f $ is bounded, it is well-known that $ \sup_{\lambda \in (-\pi,\pi]} \textrm{Var}(I_{1,b}(\lambda))=\mathcal{O}(1) $. Moreover, Theorem 2.2 of \citeasnoun{Dahlhaus85} yields $ \sup_{\lambda \in (-\pi,\pi]} \textrm{Var}(\widetilde{f}_b(\lambda))=\mathcal{O}(b/N) $. Therefore, it follows
\bee
\sup_{j_1,j_2 \in \mathcal{G}(b)} |B_{j_1,j_2,b}| \leq \sup_{\lambda \in (-\pi,\pi]} \sqrt{\textrm{Var}(\widetilde{f}_b(\lambda)) \textrm{Var}(I_{1,b}(\lambda))} = \mathcal{O}\big(\sqrt{b/N}\big) = o(b^{-1}) \,.
\eee
The last equality is guaranteed by the assumption $ b^3/n \rightarrow 0 $. Analogously, one can show $ \sup_{j_1,j_2 \in \mathcal{G}(b)} |C_{j_1,j_2,b}| = o(b^{-1}) $. Using boundedness of the function $ \varphi $, $ |\mathcal{G}(b)|=\mathcal{O}(b) $, and the obtained results for $ B_{j_1,j_2,b} $ and $ C_{j_1,j_2,b} $ it is apparent from \eqref{MMlem21} that
\be
E(W_{1,b}^2)=\frac{(2\pi)^2}{b} \sum_{j_1,j_2 \in \mathcal{G}(b)} \varphi(\lambda_{j_1,b}) \varphi(\lambda_{j_2,b}) \, A_{j_1,j_2,b} + o(1)= R1+R2 + o(1) \,, \label{MMlem22}
\ee
where
\begin{align}
R1 &:= \frac{(2\pi)^2}{b} \sum_{j \in \mathcal{G}(b)} \varphi(\lambda_{j,b}) \big( \varphi(\lambda_{j,b}) + \varphi(-\lambda_{j,b}) \big) \, \textrm{Var}(I_{1,b}(\lambda_{j,b})) \,, \nonumber\\
R2 &:= \frac{(2\pi)^2}{b} \sum_{j_1 \in \mathcal{G}(b)} \sum_{j_2 \in \mathcal{G}(b)\setminus \{j_1,-j_1\}} \varphi(\lambda_{j_1,b}) \, \varphi(\lambda_{j_2,b}) \, \textrm{Cov}(I_{1,b}(\lambda_{j_1,b}),I_{1,b}(\lambda_{j_2,b})) \,. \label{R1R2}
\end{align}
As for $ R1 $ and $ R2 $, note that $ I_{1,b}(\lambda_{j_1,b})=I_{1,b}(\lambda_{j_2,b}) $ whenever $ j_2=j_1 $ or $ j_2 = -j_1 $, and that $ \lambda_{-j,b}=-\lambda_{j,b} $. According to \citeasnoun{Krogstad82}, specifically equation (4.7), Proposition 4.6 $(ii)$ and Corollary 4.5 therein, we get
\be
& & \textrm{Cov}\big( I_{1,b}(\lambda_{j_1,b}), I_{1,b}(\lambda_{j_2,b}) \big) \nonumber \\
&=& \begin{cases}
(1+\mathds{1}_{\{|\lambda_{j_1,b}|=\pi \}})f(\lambda_{j_1,b})^2 + o(1), & \textrm{ if } |j_1|=|j_2|\, , \\
\frac{2\pi}{b} \, f_4(\lambda_{j_1,b},\lambda_{j_2,b},-\lambda_{j_2,b}) \, (1+o(1)) + \mathcal{O}(b^{-2}), & \textrm{ if } |j_1|\neq |j_2| \, ,
\end{cases} \label{MMlem23}
\ee
where the $ \mathcal{O} $ and $ o $ terms are uniform over all eligible frequencies. Since it holds $ |\mathcal{G}(b)|=\mathcal{O}(b) $ it follows from \eqref{MMlem23}
\bee
R1 &=&  2\pi \, \sum_{j \in \mathcal{G}(b)} \frac{2\pi}{b} \, \varphi(\lambda_{j,b}) \big( \varphi(\lambda_{j,b}) + \varphi(-\lambda_{j,b}) \big) \, f(\lambda_{j,b})^2 + o(1) \\
&=& 2\pi \, \int_{-\pi}^{\pi} \varphi(\lambda) \big( \varphi(\lambda) + \varphi(-\lambda) \big) \, f(\lambda)^2 \, d\lambda + o(1),
\eee
because the last sum expression is a Riemann sum due to $ \lambda_{j,b}=2\pi j/b $. This yields the first part of the limiting variance $ \tau^2 $.\\
For expression $ R2 $ we can use \eqref{MMlem23} to derive
\bee
R2 &=& \frac{(2\pi)^2}{b} \sum_{j_1 \in \mathcal{G}(b)} \sum_{j_2 \in \mathcal{G}(b)\setminus \{j_1,-j_1\}} \varphi(\lambda_{j_1,b}) \, \varphi(\lambda_{j_2,b}) \, \frac{2\pi}{b} \, f_4(\lambda_{j_1,b},\lambda_{j_2,b},-\lambda_{j_2,b}) \, (1+o(1)) \\
& & + \frac{(2\pi)^2}{b} \sum_{j_1 \in \mathcal{G}(b)} \sum_{j_2 \in \mathcal{G}(b)\setminus \{j_1,-j_1\}} \varphi(\lambda_{j_1,b}) \, \varphi(\lambda_{j_2,b}) \, \mathcal{O}(b^{-2}) +o(1) \\
&=& 2\pi \, \sum_{j_1,j_2 \in \mathcal{G}(b)}  \frac{(2\pi)^2}{b^2} \, \varphi(\lambda_{j_1,b}) \, \varphi(\lambda_{j_2,b}) \, f_4(\lambda_{j_1,b},\lambda_{j_2,b},-\lambda_{j_2,b}) + o(1) + \mathcal{O}(b^{-1}) \\
&=& 2\pi \, \int_{-\pi}^{\pi} \int_{-\pi}^{\pi} \varphi(\lambda_1) \, \varphi(\lambda_2) \, f_4(\lambda_1,\lambda_2,-\lambda_2) \, d\lambda_1 \, d\lambda_2 + o(1),
\eee
since the last sum expression is a Riemann sum. Combining the results for $ R1 $ and $ R2 $, this completes the proof. $ \hfill \square $

\

{\sc Proof of Lemma \ref{lemma2} $ (ii) $:} We have $ b=b(n) \rightarrow \infty $, thus \eqref{CLT} yields
\bee
L_b=\sqrt{b} \int_{-\pi}^{\pi} \varphi(\lambda) \, (I_{1,b}(\lambda)-f(\lambda)) \, d\lambda  \stackrel{d}{\longrightarrow} \mathcal{N}(0,\tau^2) \,.
\eee
Since $ \varphi $ is of bounded variation and $ f $ is bounded, Theorem~5.10.2~of \citeasnoun{Brillinger} then implies that the (at the Fourier frequencies) discretized version of $ L_b $ has the same limiting distribution, that is,
\be
D_b:=\sqrt{b} \sum_{j \in \mathcal{G}(b)} \frac{2\pi}{b} \, \varphi(\lambda_{j,b}) \, (I_{1,b}(\lambda_{j,b})-f(\lambda_{j,b}))  \stackrel{d}{\longrightarrow} \mathcal{N}(0,\tau^2) \,. \label{MMlem2h2}
\ee
Note that the result from \citeasnoun{Brillinger} can be applied here, because an inspection of its proof shows that the standing assumption of finite moments of all orders is not necessary and the assumptions in this paper are sufficient. Now consider the difference between $ D_b $ and $ W_{1,b} $:
\be
| D_b-W_{1,b} | &\leq & \frac{2\pi}{\sqrt{b}} \sup_{\lambda \in [-\pi,\pi]} |\varphi(\lambda)| \sum_{j \in \mathcal{G}(b)} |\widetilde{f}_b(\lambda_{j,b})-f(\lambda_{j,b})| \,. \label{MMlem2h1}
\ee
For the bias of $ \widetilde{f}_b $ we have from strict stationarity of $ (X_t)_{t \in \Z} $
\bee
\sup_{\lambda \in [-\pi,\pi]} |E(\widetilde{f}_b(\lambda))-f(\lambda)| = \sup_{\lambda \in [-\pi,\pi]} |E(I_{1,b}(\lambda))-f(\lambda)| = \mathcal{O}(b^{-1})\,,
\eee
the last equation is -- under the conditions of Assumption \ref{assu_1} -- a well-known result. This can be used together with Lemma \ref{lemma1} $ (i) $ to obtain
\begin{align} \label{eq.ftilde}
\sum_{j \in \mathcal{G}(b)} |\widetilde{f}_b(\lambda_{j,b})-f(\lambda_{j,b})| &\leq  \sum_{j \in \mathcal{G}(b)} |\widetilde{f}_b(\lambda_{j,b}) -E(\widetilde{f}_b(\lambda_{j,b}))| + b \,\sup_{\lambda \in [-\pi,\pi]} |E(\widetilde{f}_b(\lambda))-f(\lambda)| \ \nonumber \\
&= \mathcal{O}_P \big(\sqrt{b^3/N}\big) + \mathcal{O}(1) = \mathcal{O}_P(1),
\end{align}
due to $ b^3/n =o(1) $. Using this bound in \eqref{MMlem2h1}, and boundedness of $ \varphi $, yields $ | D_b-W_{1,b} |=o_P(b^{-1/2}) $. Combined with \eqref{MMlem2h2}, this completes the proof. $ \hfill \square $

\

{\sc Proof of Proposition \ref{prop1} $ (i) $:} Since the $ U_j^* $ are for $ j=1,\ldots,[n/2] $ i.i.d.~standard exponentially distributed, and for $ j<0 $ it holds $ U_j^*=U_{-j}^* $, we have $ {\rm Cov}^*(U_{j_1}^*,U_{j_2}^*)=\mathds{1}_{\{|j_1|=|j_2|\}} $. Hence, with $ \widehat{f}_n $ being an even function, it follows\\$ {\rm Cov}^*( T^*(\lambda_{j_1,n}),T^*(\lambda_{j_2,n}))= \widehat{f}_n(\lambda_{j_1,n})^2 \, \mathds{1}_{\{|j_1|=|j_2|\}} $ and
\bee
{\rm Var}^*(V_{n}^*) &=& \frac{4\pi^2}{n} \sum_{j_1,j_2 \in \mathcal{G}(n)} \varphi(\lambda_{j_1,n}) \, \varphi(\lambda_{j_2,n}) \, {\rm Cov}^*( T^*(\lambda_{j_1,n}),T^*(\lambda_{j_2,n}) ) \\
&=& 2\pi \, \sum_{j \in \mathcal{G}(n)} \frac{2\pi}{n} \, \varphi(\lambda_{j,n}) \,\big( \varphi(\lambda_{j,n}) + \varphi(-\lambda_{j,n}) \big) \, f(\lambda_{j,n})^2 \\
& & + \frac{4\pi^2}{n} \sum_{j \in \mathcal{G}(n)} \varphi(\lambda_{j,n}) \,\big( \varphi(\lambda_{j,n}) + \varphi(-\lambda_{j,n}) \big) \, \big(\widehat{f}_n(\lambda_{j,n})^2 - f(\lambda_{j,n})^2 \big) \\
&=:& S_1 + S_2 \,.
\eee
Since $ S_1 $ is a Riemann sum, it follows $ S_1=\tau_1^2+o(1) $. Using $ |\mathcal{G}(n)|=\mathcal{O}(n) $ and Assumption \ref{assu_fnhat}, it is easy to see that $ S_2=o_P(1) $ which completes the proof of $ (i) $. $ \hfill \square $

\

{\sc Proof of Proposition \ref{prop1} $ (ii) $:} Since the $ U_j^* $ are for $ j=1,\ldots,[n/2] $ i.i.d.~standard exponentially distributed, and for $ j<0 $ it holds $ U_j^*=U_{-j}^* $, we define $ \mathcal{G}(n,+):=\{ 1,\ldots, [n/2] \} $, the positive part of $ \mathcal{G}(n) $. Then we can write $ V_{n}^* $ as a sum of (conditionally) independent mean zero random variables according to
\be
V_{n}^*=\sum_{j \in \mathcal{G}(n,+)} (Z_{j,n}^*/\sqrt{n}) \,, \label{MMtheo1h5}
\ee
where
\bee
Z_{j,n}^* := 2\pi \, (U_j^*-1) \, \big( \varphi(\lambda_{j,n}) \, \widehat{f}_n(\lambda_{j,n}) + \varphi(-\lambda_{j,n}) \, \widehat{f}_n(-\lambda_{j,n}) \big) \,, \quad j \in \mathcal{G}(n,+) \,.
\eee
Since $ |\mathcal{G}(n,+)| \rightarrow \infty $, as $ n \rightarrow \infty $, a conditional version of Lyapunov's CLT can be applied to \eqref{MMtheo1h5}. We verify Lyapunov's condition which in this case (using the established convergence in probability of $ {\rm Var}^*(V_{n}^*) $ to a positive constant from $ (i) $) means we have to find some $ \delta > 0 $ such that
\be
n^{-(1+\delta/2)} \, \sum_{j \in \mathcal{G}(n,+)} E^*\left(|Z_{j,n}^*|^{2+\delta}\right) = o_P(1) \,. \label{MMtheo1h7}
\ee
We choose $ \delta=1 $ and proceed in the following way: First, as a preliminary, we have
\bee
& & \sum_{j \in \mathcal{G}(n,+)} \big| \varphi(\lambda_{j,n}) \, \widehat{f}_n(\lambda_{j,n}) + \varphi(-\lambda_{j,n}) \, \widehat{f}_n(-\lambda_{j,n}) \big| \\
&\leq & \sup_{\lambda \in [-\pi,\pi]} |\varphi(\lambda)| \cdot n \cdot \sup_{\lambda \in [-\pi,\pi]} |\widehat{f}_n(\lambda)| = \mathcal{O}_P(n), \\
\eee
because by Assumption \ref{assu_1} both $ \varphi $ and the spectral density $ f $ are bounded ($ f $ is bounded by $ \sum_{h \in \Z} |\gamma(h)|<\infty $), and Assumption \ref{assu_fnhat} guarantees $ \sup_{\lambda \in [-\pi,\pi]} |\widehat{f}_n(\lambda)-f(\lambda)| = o_P(1) $. One can now follow along these lines to obtain
\be
\sum_{j \in \mathcal{G}(n,+)} \big| \varphi(\lambda_{j,n}) \, \widehat{f}_n(\lambda_{j,n}) + \varphi(-\lambda_{j,n}) \, \widehat{f}_n(-\lambda_{j,n}) \big|^3 = \mathcal{O}_P(n)\,. \label{MMtheo1h6}
\ee
Then, a straightforward calculation yields $ E^*(|U_j^*-1|^3)=12e^{-1}-2 $ because all $ U_j^* $ are standard exponential. This implies
\bee
& & \sum_{j \in \mathcal{G}(n,+)} E^*\left(|Z_{j,n}^*|^3 \right) \\
&=& 8\pi^3(12e^{-1}-2) \, \sum_{j \in \mathcal{G}(n,+)} \big| \varphi(\lambda_{j,n}) \, \widehat{f}_n(\lambda_{j,n}) + \varphi(-\lambda_{j,n}) \, \widehat{f}_n(-\lambda_{j,n}) \big|^3 = \mathcal{O}_P(n)\,,
\eee
due to \eqref{MMtheo1h6}. This shows that the Lyapunov condition \eqref{MMtheo1h7} holds for $ \delta=1 $. Together with the result for the variance from $ (i) $, the assertion follows. $ \hfill \square $.

\

{\sc Proof of Theorem \ref{theorem1} $ (i) $:} For each Fourier frequency $ \lambda_{j,b} $, $ j \in \mathcal{G}(b) $, the random variables $ I_b^{(1)}(\lambda_{j,b}),\ldots,I_b^{(k)}(\lambda_{j,b}) $ are conditionally independent and each possess a discrete uniform distribution on the set
\bee
\left\{ \frac{\widehat{f}_n(\lambda_{j,b})}{\widetilde{f}_b(\lambda_{j,b})} \,I_{1,b}(\lambda_{j,b}), \, \ldots \, , \frac{\widehat{f}_n(\lambda_{j,b})}{\widetilde{f}_b(\lambda_{j,b})} \,I_{N,b}(\lambda_{j,b}) \right\} \,.
\eee
Hence, we get as a preliminary consideration:
\be
& & \textrm{Cov}^*\left(I_b^{(1)}(\lambda_{j_1,b}),I_b^{(1)}(\lambda_{j_2,b})\right) \nonumber \\
&=& E^*\left(I_b^{(1)}(\lambda_{j_1,b}) \, I_b^{(1)}(\lambda_{j_2,b})\right) - E^*(I_b^{(1)}(\lambda_{j_1,b})) \, E^*( I_b^{(1)}(\lambda_{j_2,b})) \nonumber \\
&=& \frac{\widehat{f}_n(\lambda_{j_1,b})\widehat{f}_n(\lambda_{j_2,b})}{\widetilde{f}_b(\lambda_{j_1,b})\widetilde{f}_b(\lambda_{j_2,b})} \, \left[ \frac{1}{N} \sum_{t=1}^{N} I_{t,b}(\lambda_{j_1,b}) \, I_{t,b}(\lambda_{j_2,b}) - \widetilde{f}_b(\lambda_{j_1,b}) \, \widetilde{f}_b(\lambda_{j_2,b}) \right] \nonumber\\
&=:& Q_{j_1,j_2,n} \cdot H_{j_1,j_2,n} \,. \label{MMtheo1h2}
\ee
Next, note that due to the assumption $ b^3/n =o(1) $ the expression appearing in Lemma \ref{lemma1} $(ii)$ vanishes with rate $ o_P(b) $, which will be used in the ensuing calculation. Lemma \ref{lemma1} now yields
\be
& & \sum_{j_1,j_2 \in \mathcal{G}(b)} \big| H_{j_1,j_2,n} -\textrm{Cov}\left(I_{1,b}(\lambda_{j_1,b}),I_{1,b}(\lambda_{j_2,b})\right) \big| \nonumber \\
&\leq & \sum_{j_1,j_2 \in \mathcal{G}(b)} \Big| \frac{1}{N} \sum_{t=1}^{N} I_{t,b}(\lambda_{j_1,b}) \, I_{t,b}(\lambda_{j_2,b}) - E\left(I_{1,b}(\lambda_{j_1,b}) \, I_{1,b}(\lambda_{j_2,b})\right) \Big| \nonumber \\
& & + \sum_{j_1,j_2 \in \mathcal{G}(b)} \Big| \widetilde{f}_b(\lambda_{j_1,b}) \, \widetilde{f}_b(\lambda_{j_2,b}) - EI_{1,b}(\lambda_{j_1,b}) \, EI_{1,b}(\lambda_{j_2,b}) \Big| \nonumber \\
&=& o_P(b) \,, \label{MMtheo1h1}
\ee
the bound for the second sum expression can be deduced from Lemma \ref{lemma1} $(i)$ using $ \sum_{j \in \mathcal{G}(b)} |\widetilde{f}_b(\lambda_{j,b})|= \mathcal{O}_P(b) $ and $ \sum_{j \in \mathcal{G}(b)} |EI_{1,b}(\lambda_{j,b})|= \mathcal{O}(b) $. The previous calculation \eqref{MMtheo1h1} together with \eqref{MMlem23} can be used to derive
\be
\sum_{j_1,j_2 \in \mathcal{G}(b)} | H_{j_1,j_2,n} | &\leq & \sum_{j_1,j_2 \in \mathcal{G}(b)} \big| \textrm{Cov}\left(I_{1,b}(\lambda_{j_1,b}),I_{1,b}(\lambda_{j_2,b})\right) \big| \nonumber\\
& & + \sum_{j_1,j_2 \in \mathcal{G}(b)} \big| H_{j_1,j_2,n} -\textrm{Cov}\left(I_{1,b}(\lambda_{j_1,b}),I_{1,b}(\lambda_{j_2,b})\right) \big| \nonumber \\
&\leq & \mathcal{O}(b) + o_P(b) = \mathcal{O}_P(b) \,. \label{MMtheo1h10}
\ee
Moreover, notice that by Assumption 3 and \eqref{eq.ftilde} -- which implies $ \sup_{j\in {\mathcal G}(b)} | \widetilde{f}_b(\lambda_{j,b}) - f(\lambda_{j,b}) | =o_P(1) $ -- as well as the assumption that $f(\lambda)$ is bounded away from zero in the interval $ [-\pi,\pi]$, we get that
\begin{equation} \label{eq.unif-f}
\sup_{j\in {\mathcal G}(b)} \Big|\frac{\widehat{f}_n(\lambda_{j,b})}{\widetilde{f}_b(\lambda_{j,b})} -1 \Big| \stackrel{P}{\rightarrow} 0 \quad \mbox{and} \quad
\sup_{j_1,j_2\in {\mathcal G}(b)} \Big|\frac{\widehat{f}_n(\lambda_{j_1,b})\widehat{f}_n(\lambda_{j_2,b})}{\widetilde{f}_b(\lambda_{j_1,b})\widetilde{f}_b(\lambda_{j_2,b})} -1 \Big| \stackrel{P}{\rightarrow} 0 \,.
\end{equation}
The results from \eqref{MMtheo1h2} through \eqref{eq.unif-f} can be combined to replace bootstrap covariances by their respective non-bootstrap counterparts via
\be
& & \sum_{j_1,j_2 \in \mathcal{G}(b)} \big| \textrm{Cov}^*\big(I_b^{(1)}(\lambda_{j_1,b}),I_b^{(1)}(\lambda_{j_2,b})\big) -\textrm{Cov}\left(I_{1,b}(\lambda_{j_1,b}),I_{1,b}(\lambda_{j_2,b})\right) \big| \nonumber \\
&\leq & \sup_{j_1,j_2\in {\mathcal G}(b)} |Q_{j_1,j_2,n}-1| \, \sum_{j_1,j_2 \in \mathcal{G}(b)} | H_{j_1,j_2,n} | \nonumber \\
& & + \sum_{j_1,j_2 \in \mathcal{G}(b)} \big| H_{j_1,j_2,n} -\textrm{Cov}\left(I_{1,b}(\lambda_{j_1,b}),I_{1,b}(\lambda_{j_2,b})\right) \big| \nonumber \\
&=& o_P(b) \,. \label{MMtheo1h1a}
\ee
After these preliminaries, using conditional independence of $ I_b^{(l_1)}(\lambda_{j_1,b}) $ and $ I_b^{(l_2)}(\lambda_{j_2,b}) $ for $ l_1 \neq l_2 $, we get
\begin{align}
\textrm{Var}^*(L_{n}^*) &= \frac{(2\pi)^2n}{b^2} \sum_{j_1,j_2 \in \mathcal{G}(b)} \varphi(\lambda_{j_1,b}) \, \varphi(\lambda_{j_2,b}) \, \textrm{Cov}^*(I_{j_1,b}^*,I_{j_2,b}^*) \nonumber \\
&= \frac{(2\pi)^2n}{b^2} \sum_{j_1,j_2 \in \mathcal{G}(b)} \varphi(\lambda_{j_1,b}) \, \varphi(\lambda_{j_2,b}) \, \frac{1}{k} \textrm{Cov}^*(I_b^{(1)}(\lambda_{j_1,b}),I_b^{(1)}(\lambda_{j_2,b})) \nonumber \\
&= R1^* + R2^* \,, \nonumber
\end{align}
where $R_1^*$ and $ R_2^*$ are defined as
\begin{align}
R1^* &:= \frac{(2\pi)^2}{b} \sum_{j \in \mathcal{G}(b)} \varphi(\lambda_{j,b}) \big( \varphi(\lambda_{j,b}) + \varphi(-\lambda_{j,b}) \big) \, \textrm{Var}^*(I_b^{(1)}(\lambda_{j,b})) \,, \nonumber\\
R2^* &:= \frac{(2\pi)^2}{b} \sum_{j_1 \in \mathcal{G}(b)} \sum_{j_2 \in \mathcal{G}(b)\setminus \{j_1,-j_1\}} \varphi(\lambda_{j_1,b}) \, \varphi(\lambda_{j_2,b}) \, \textrm{Cov}^*(I_b^{(1)}(\lambda_{j_1,b}),I_b^{(1)}(\lambda_{j_2,b})) \,. \label{R1R2star}
\end{align}
As for $ R1^* $ and $ R2^* $, note that $ I_b^{(1)}(\lambda_{j_1,b})=I_b^{(1)}(\lambda_{j_2,b}) $ whenever $ j_2=j_1 $ or $ j_2 = -j_1 $, and that $ \lambda_{-j,b}=-\lambda_{j,b} $. $ R1^* $ and $ R2^* $ are bootstrap analogues of $ R1 $ and $ R2 $ from \eqref{R1R2}. Now, using \eqref{MMtheo1h1a}, $ |\mathcal{G}(b)|=\mathcal{O}(b) $, and the boundedness of $ \varphi $ it follows that $ |R1^*-R1| $ and $ |R2^*-R2| $ vanish asymptotically in probability. Moreover, invoking the limiting results for $ R1 $ and $ R2 $ from Lemma \ref{lemma2} $ (i) $, we have
\be
R1^* &=& R1 + o_P(1) = \tau_1^2 + o_P(1)\,,\nonumber \\
R2^* &=& R2 + o_P(1) = \tau_2 + o_P(1)\,. \label{MMlimR1R2star}
\ee
With $ \tau^2=\tau_1^2+\tau_2 $, this completes the proof of $ (i) $. $ \hfill \square $

\

{\sc Proof of Theorem \ref{theorem1} $ (ii) $:} Since $ L_n $ is asymptotically normal according to \eqref{CLT}, which implies $ \sup_{x \in \R} | P(L_n \leq x) - \Phi(x/\tau) | =o(1) $, it suffices to show
\bee
\sup_{x \in \R} | P^*(L_{n}^* \leq x) - \Phi(x/\tau) | =o_P(1) \,.
\eee
Notice that by definition of $ L_n^* $, and recalling that $ i_1^*,\ldots,i_k^* $ are (conditionally) i.i.d.~with a discrete uniform distribution on $ \{1,\ldots,N\} $, we have
\begin{align*}
L_n^* & = \frac{1}{\sqrt{k}}\sum_{l=1}^k \frac{2\pi}{\sqrt{b}} \sum_{j \in \mathcal{G}(b)} \varphi(\lambda_{j,b})  \left( I_{i_l^*,b}(\lambda_{j,b}) - \widetilde{f}_b(\lambda_{j,b}) \right)\\
& \ \ \ \ +  \frac{1}{\sqrt{k}}\sum_{l=1}^k \frac{2\pi}{\sqrt{b}} \sum_{j \in \mathcal{G}(b)} \varphi(\lambda_{j,b})\Big(\frac{\widehat{f}_n(\lambda_{j,b})}{\widetilde{f}_b(\lambda_{j,b})} -1 \Big)  \left( I_{i_l^*,b}(\lambda_{j,b}) - \widetilde{f}_b(\lambda_{j,b}) \right)\\
& =: M1^*+M2^*\,.
\end{align*}
The strategy is to first show that $ M2^* $ vanishes asymptotically in probability, and then to show that $ M1^* $ is asymptotically normal with mean zero and with the proper variance $ \tau^2 $.\\
As for $ M2^* $, we have due to (conditional) independence of $ i_l^* $ and $ i_m^* $ for all $ l\neq m $, and due to \eqref{MMtheo1h2},
\begin{align*}
\textrm{Var}^*(M2^*) &= \frac{1}{k} \sum_{l,m=1}^k \frac{4\pi^2}{b} \sum_{j_1,j_2 \in \mathcal{G}(b)} \varphi(\lambda_{j_1,b}) \varphi(\lambda_{j_2,b}) \Big( \frac{\widehat{f}_n(\lambda_{j_1,b})}{\widetilde{f}_b(\lambda_{j_1,b})} -1 \Big) \Big( \frac{\widehat{f}_n(\lambda_{j_2,b})}{\widetilde{f}_b(\lambda_{j_2,b})} -1 \Big) \\
& \quad \quad \quad \quad \quad \quad \quad \quad \quad \quad \quad \times \; \textrm{Cov}^*\left(I_{i_l^*,b}(\lambda_{j_1,b}),I_{i_m^*,b}(\lambda_{j_2,b})\right) \\
&= \frac{4\pi^2}{b} \sum_{j_1,j_2 \in \mathcal{G}(b)} \varphi(\lambda_{j_1,b}) \varphi(\lambda_{j_2,b}) \Big( \frac{\widehat{f}_n(\lambda_{j_1,b})}{\widetilde{f}_b(\lambda_{j_1,b})} -1 \Big) \Big( \frac{\widehat{f}_n(\lambda_{j_2,b})}{\widetilde{f}_b(\lambda_{j_2,b})} -1 \Big) \, H_{j_1,j_2,n} \\
&\leq \frac{4\pi^2}{b} \Big(\sup_{\lambda \in [-\pi,\pi]} |\varphi(\lambda)|\Big)^2 \left(\sup_{j \in \mathcal{G}(b)}\Big| \frac{\widehat{f}_n(\lambda_{j,b})}{\widetilde{f}_b(\lambda_{j,b})} -1 \Big| \right)^2 \sum_{j_1,j_2 \in \mathcal{G}(b)} | H_{j_1,j_2,n} | \\
&= o_P(1)\,,
\end{align*}
where boundedness of $ \varphi $ and assertions \eqref{MMtheo1h10} and \eqref{eq.unif-f} have been used. This is sufficient for
\bee
P^*(|M2^*| > \varepsilon ) =o_P(1), \quad \forall \varepsilon > 0\,,
\eee
that is, $ M2^* $ vanishes asymptotically in probability. As for $ M1^* $, we can write
\bee
M1^* = \sum_{l=1}^{k} W_b^{(l)}/\sqrt{k}\,,
\eee
with
\bee
W_b^{(l)} := \frac{2\pi}{\sqrt{b}} \sum_{j \in \mathcal{G}(b)} \varphi(\lambda_{j,b}) \left( I_{i_l^*,b}(\lambda_{j,b}) - \widetilde{f}_b(\lambda_{j,b}) \right) \,,
\eee
and $ W_b^{(1)},\ldots,W_b^{(k)} $ are -- for each $ n \in \N $ and conditionally on the original data sample $ X_1,\ldots,X_n $ -- i.i.d.~random variables with a discrete uniform distribution on $ \{ W_{1,b},\ldots,W_{N,b} \} $, as defined in \eqref{MMWtb}. Hence, $ (W_b^{(1)},\ldots,W_b^{(k)})_{n \in \N} $ form a triangular array, and a conditional version of the Lindeberg--Feller CLT can be applied. Since we already know that $ L_n^* $ and $ M1^* $ have the same limiting distribution, assertion $ (i) $ of Theorem \ref{theorem1} yields that $ \sum_{l=1}^{k} \textrm{Var}^*(W_b^{(l)}/\sqrt{k}) $ converges to $ \tau^2 $ in probability. Hence, it suffices to prove
\be
\sum_{l=1}^{k} E^*\left( \big(W_b^{(l)}/\sqrt{k} \big)^2 \, \mathds{1}_{\{ |W_b^{(l)}| \geq \varepsilon \sqrt{k} \}} \right) =o_P(1) \,, \quad \forall \varepsilon > 0\,, \label{MMtheo1h3}
\ee
in order for Lindeberg's condition to be fulfilled and, therefore, the assertion to hold. In order do this, we will first show that the sequence $ (W_{1,b}^2)_{n \in \N} $ is uniformly integrable in the sense of \citeasnoun{Billingsley95}, i.e. that
\be
\lim_{\alpha \rightarrow \infty} \sup_{n \in \N} E\left( W_{1,b}^2 \, \mathds{1}_{\{ W_{1,b}^2 \geq \alpha^2 \}} \right)=0\,, \label{MMtheo1h8}
\ee
When dealing with $ (W_{1,b}^2)_{n \in \N} $ in the following, recall that $ b=b(n) $ which is suppressed in the notation for convenience reasons. From Lemma \ref{lemma2} $ (ii) $ we already have $ W_{1,b}^2 \stackrel{d}{\rightarrow} Z^2 $, as $ n \rightarrow \infty $, with $ Z \sim \mathcal{N}(0,\tau^2) $. According to Theorem~25.6 of \citeasnoun{Billingsley95}, there exists a probability space with random variables $ (Y_n)_{n \in \N} $ and $ Y $, with $ Y \sim Z^2 $ and $ Y_n \sim W_{1,b}^2 $ for all $ n \in \N $, such that $ Y_n \rightarrow Y $ almost surely. Moreover, Lemma \ref{lemma2} $ (i) $ implies
\bee
E(Y_n)= E(W_{1,b}^2) \longrightarrow \tau^2 = E(Z^2) =E(Y)\,.
\eee
Theorem~16.14~$ (ii) $ of \citeasnoun{Billingsley95} states that this convergence of expectations together with almost sure convergence of $ Y_n $ yields uniform integrability of the non-negative sequence $ (Y_n)_{n \in \N} $, that is
\bee
\lim_{\alpha \rightarrow \infty} \sup_{n \in \N} E\left( Y_n \, \mathds{1}_{\{ Y_n \geq \alpha^2 \}} \right)=0\,.
\eee
Due to $ Y_n \sim W_{1,b}^2 $, this assertion immediately gives \eqref{MMtheo1h8}.\\
Now, we turn back to the Lindeberg condition. Since $ \sqrt{k} \rightarrow \infty $, as $ n \rightarrow \infty $, it can be verified in a straightforward way that \eqref{MMtheo1h8} implies
\be
E\big( W_{1,b}^2 \, \mathds{1}_{\{ |W_{1,b}| \geq \varepsilon \sqrt{k} \}} \big)=o(1), \quad \forall \varepsilon > 0\,. \label{MMtheo1h4}
\ee
It follows for arbitrary $ \varepsilon > 0 $, due to strict stationarity of $ (X_t)_{t \in \Z} $:
\be
& & E\left(\sum_{l=1}^{k} E^*\Big( \big(W_b^{(l)}/\sqrt{k} \big)^2 \, \mathds{1}_{\{ |W_b^{(l)}| \geq \varepsilon \sqrt{k} \}} \Big) \right) \nonumber\\
&=& E\left( E^*\big( (W_b^{(1)})^2 \, \mathds{1}_{\{ |W_b^{(1)}| \geq \varepsilon \sqrt{k} \}} \big) \right) \nonumber\\
&=& E\left( \frac{1}{N} \sum_{t=1}^{N} W_{t,b}^2 \, \mathds{1}_{\{ |W_{t,b}| \geq \varepsilon \sqrt{k} \}} \right) = E\left( W_{1,b}^2 \, \mathds{1}_{\{ |W_{1,b}| \geq \varepsilon \sqrt{k} \}} \right) =o(1) \,, \label{MMtheo1h9}
\ee
due to \eqref{MMtheo1h4}. Assertion \eqref{MMtheo1h9} is sufficient for \eqref{MMtheo1h3} which completes the proof of $ (ii) $. $ \hfill \square $

\

{\sc Proof of Theorem \ref{theorem1} $ (iii) $:} Comparing the definitions of $ L_{n,R}^* $ and $ V_{2,n}^* $ from \eqref{V2star-Ratio} and \eqref{eq.diffVs}, respectively, and due to the structure of the function $ \widetilde{w} $ defined in \eqref{wtilde}, $ L_{n,R}^* $ can be written as
\bee
L_{n,R}^* = \frac{1}{\left( \frac{2\pi}{b} \sum_{j \in \mathcal{G}(b)} I_{j,b}^*\right) \left( \frac{2\pi}{b} \sum_{j \in \mathcal{G}(b)} \widehat{f}_n(\lambda_{j,b})\right) } \, V_{2,n}^* =: \frac{1}{G_n^*} V_{2,n}^* \,.
\eee
In view of \eqref{CLT-Ratio} it suffices to show $ \sup_{x \in \R} | P^*( L_{n,R}^*\leq x ) - \Phi(x/\tau_{R})|=o_P(1) $, where $ \Phi $ denotes the cdf of the standard normal distribution. Invoking a conditional version of Slutsky's Lemma, this will be established by showing that
\begin{enumerate}
\item[(a)] \ $ \textrm{Var}^*(V_{2,n}^*) \stackrel{P}{\rightarrow} M(1,f)^4\, \tau^2_{R}$,
\item[(b)] \ $ \sup_{x \in \R} | P^*( V_{2,n}^*\leq x ) - \Phi(x/[M(1,f)^2\, \tau_{R}])|=o_P(1) $,
\item[(c)] \ $ P^*(|G_n^* - M(1,f)^2|> \varepsilon) =o_P(1) \quad \quad \forall \, \varepsilon >0 $.
\end{enumerate}
Verify first that from Assumption \ref{assu_fnhat} we have $\sup_{\lambda\in [-\pi,\pi]}|\widetilde{w}(\lambda)-w(\lambda)| \stackrel{P}{\rightarrow} 0$. Moreover, the form of $ \widetilde{w} $ yields $ 2\pi b^{-1} \sum_{j \in \mathcal{G}(b)} \widetilde{w}(\lambda_{j,b}) \widehat{f}_n(\lambda_{j,b})=0 $. Hence,
\be
V_{2,n}^* = \sqrt{n} \,\frac{2\pi}{b} \sum_{j \in \mathcal{G}(b)} \widetilde{w}(\lambda_{j,b}) \big( I_{j,b}^* - \widehat{f}_n(\lambda_{j,b}) \big) \,, \label{MMtheo2h2}
\ee
which means that $ V_{2,n}^* $ equals $ L_n^* $ if one replaces $ \varphi $ by its linear transformation $ \widetilde{w} $. Therefore, assertion (a) follows with the same arguments as the result for $ \textrm{Var}^*(L_n^*) $ in the proof of Theorem \ref{theorem1} $ (i) $, after replacing $ \widetilde{w} $ by $ w $ via $\sup_{\lambda\in [-\pi,\pi]}|\widetilde{w}(\lambda)-w(\lambda)| \stackrel{P}{\rightarrow} 0$.\\
As for assertion (c), Assumption \ref{assu_fnhat} can be invoked to obtain
\be
E^*\Big( \frac{2\pi}{b} \sum_{j \in \mathcal{G}(b)} I_{j,b}^* \Big) &=& \frac{2\pi}{b} \sum_{j \in \mathcal{G}(b)} \widehat{f}_n(\lambda_{j,b}) = \frac{2\pi}{b} \sum_{j \in \mathcal{G}(b)} f(\lambda_{j,b}) + o_P(1) \nonumber\\
&=& M(1,f) + o_P(1)\,. \label{MMtheo2h1}
\ee
Also, $ \textrm{Var}^*( 2\pi b^{-1} \sum_{j \in \mathcal{G}(b)} I_{j,b}^* )=n^{-1}\textrm{Var}^*(L_n^*) $ if one uses the function $ \varphi(\lambda)=1 $ in $ L_n^* $. Therefore, $ \textrm{Var}^*( 2\pi b^{-1} \sum_{j \in \mathcal{G}(b)} I_{j,b}^* ) =\mathcal{O}_P(n^{-1}) $ by Theorem \ref{theorem1} $ (i) $, which is sufficient for
\bee
P^*\Big(\Big|\frac{2\pi}{b} \sum_{j \in \mathcal{G}(b)} I_{j,b}^* - M(1,f) \Big|> \varepsilon \Big) =o_P(1) \quad \quad \forall \, \varepsilon >0 \,.
\eee
Since the second factor in $ G_n^* $ also converges to $ M(1,f) $ in probability due to \eqref{MMtheo2h1}, assertion (c) holds true.\\
Finally, to establish (b), note that from representation \eqref{MMtheo2h2} $ V_{2,n}^* $ can be decomposed as
\be
V_{2,n}^* &=& \frac{1}{\sqrt{k}} \sum_{l=1}^{k} \frac{2\pi}{\sqrt{b}} \sum_{j \in \mathcal{G}(b)} w(\lambda_{j,b}) \big( I_{i_l^*,b} - \widetilde{f}_n(\lambda_{j,b}) \big) \nonumber\\
& & + \frac{1}{\sqrt{k}} \sum_{l=1}^{k} \frac{2\pi}{\sqrt{b}} \sum_{j \in \mathcal{G}(b)} (\widetilde{w}(\lambda_{j,b}) - w(\lambda_{j,b})) \big( I_{i_l^*,b} - \widetilde{f}_n(\lambda_{j,b}) \big) \nonumber\\
& & + \frac{1}{\sqrt{k}} \sum_{l=1}^{k} \frac{2\pi}{\sqrt{b}} \sum_{j \in \mathcal{G}(b)} \widetilde{w}(\lambda_{j,b}) \Big( \frac{\widehat{f}_n(\lambda_{j,b})}{\widetilde{f}_b(\lambda_{j,b})} -1 \Big) \big( I_{i_l^*,b} - \widetilde{f}_n(\lambda_{j,b}) \big) \,. \label{MMtheo2h3}
\ee
The first summand on the right-hand side equals expression $ M1^* $ from the proof of Theorem \ref{theorem1} $ (ii) $ if one replaces $ \varphi $ by $ w $ there. Hence, asymptotic normality of the first summand follows along these lines (and the proper variance has already been established in (a)). The third summand on the right-hand side of \eqref{MMtheo2h3} can be treated similar to expression $ M2^* $ from the proof of Theorem \ref{theorem1} $ (ii) $, and therefore vanishes asymptotically in probability using $\sup_{\lambda\in [-\pi,\pi]}|\widetilde{w}(\lambda)-w(\lambda)| \stackrel{P}{\rightarrow} 0$. With the same argument the second summand vanishes which establishes asymptotic normality of $ V_{2,n}^* $ and completes the proof. $ \hfill \square $

\

{\sc Proof of Theorem \ref{theorem2} $ (i) $:} From the definition of $ R1^* $ in \eqref{R1R2star} and from \eqref{MMtheo1h2} it is obvious that actually $ c_n=R1^* $. Hence, \eqref{MMlimR1R2star} implies $ c_n = \tau_1^2 + o_P(1) $. In particular, the variance correction term $ c_n $ represents exactly the contribution of $ L_{n}^* $ to $ \tau_1^2 $ of the limiting variance, while $ R2^* $ delivers the contribution to $ \tau_2 $. Moreover, since $ V_{n}^* $ and $ L_{n}^* $ are conditionally independent, we get from Proposition \ref{prop1} $ (i) $ and Theorem \ref{theorem1} $ (i) $ that $ {\rm Var}^*(L^*_n+ V_{n}^*) = \tau_1^2+\tau^2 + o_P(1) $. This yields
\bee
\widehat \tau^2= {\rm Var}^*(L^*_n+ V_{n}^*)-c_n \stackrel{P}{\longrightarrow} \tau^2 \,.
\eee
Also we have $ \widehat \tau_1^2=\tau_1^2+o_P(1) $. This completes the proof of $ (i) $. $ \hfill \square $

\

{\sc Proof of Theorem \ref{theorem2} $ (ii) $:}
Heuristically, the idea is that the (conditional) distribution of $ V_{n}^* $ converges to $ \mathcal{N}(0,\tau_1^2) $ in probability, while $ \widehat \tau/\widehat \tau_1 = \tau/\tau_1 + o_P(1) $ and, therefore, the distribution of $ \widetilde{V}_n^*=(\widehat \tau/\widehat \tau_1) \, V_{n}^* $ converges in probability to $ \mathcal{N}(0,\tau^2) $. In the following we will prove this in a more rigorous way. Denote the cdf of the standard normal distribution by $ \Phi $. Observe that $ \widehat \tau=\tau+o_P(1) $ yields with a standard argument that
\bee
\sup_{x \in \R} \Big| \Phi\Big(\frac{x}{\widehat \tau}\Big) - \Phi\Big(\frac{x}{\tau}\Big) \Big| =o_P(1) \,,
\eee
the same being true when $ \widehat \tau $ and $ \tau $ are replaced by $ \widehat \tau_1 $ and $ \tau_1 $, respectively. Using this assertion twice, we can proceed by
\bee
& & \sup_{x \in \R} \Big| P^*(\widetilde{V}_n^*\leq x) - \Phi\Big(\frac{x}{\tau}\Big) \Big| \\
&\leq& \sup_{x \in \R} \Big| P^*(\widetilde{V}_n^*\leq x) - \Phi\Big(\frac{x}{\widehat\tau}\Big) \Big| + \sup_{x \in \R} \Big| \Phi\Big(\frac{x}{\widehat\tau}\Big) - \Phi\Big(\frac{x}{\tau}\Big) \Big| \\
&=& \sup_{x \in \R} \Big| P^* \Big( V_{n}^*\leq \frac{\widehat \tau_1 x}{\widehat\tau} \Big) - \Phi\Big(\frac{x}{\widehat\tau}\Big) \Big| + o_P(1) \\
&=& \sup_{x \in \R} \Big| P^* ( V_{n}^*\leq x ) - \Phi\Big(\frac{x}{\widehat \tau_1}\Big) \Big| + o_P(1) \\
&\leq & \sup_{x \in \R} \Big| P^* ( V_{n}^*\leq x ) - \Phi\Big(\frac{x}{\tau_1}\Big) \Big| +  \sup_{x \in \R} \Big| \Phi\Big(\frac{x}{\widehat \tau_1}\Big) - \Phi\Big(\frac{x}{\tau_1}\Big) \Big| + o_P(1) \\
&=& o_P(1) \,,
\eee
due to Proposition \ref{prop1} $ (ii) $. This completes the proof because \eqref{CLT} implies\\$ \sup_{x \in \R} | P ( L_n\leq x ) - \Phi(x/\tau)|=o(1) $. $ \hfill \square $

\

{\sc Proof of Theorem \ref{theorem2} $(iii)$:}  The proof is established by showing that
\begin{enumerate}
\item[(a)] \ $ \widehat{\sigma}_{1,R}^2 \stackrel{P}{\rightarrow} M(1,f)^4\cdot\tau^2_{1,R}$,
\item[(b)] \ $ \widehat{\sigma}^2_R \stackrel{P}{\rightarrow} M(1,f)^4 \cdot \tau^2_R$ , and
\item[(c)] \ $ \sup_{x \in \R} | P^*( V_{n,R}^*\leq x ) - \Phi(x/\tau_{1,R})|=o_P(1) $.
\end{enumerate}
Verify first that  from Assumption \ref{assu_fnhat} we have that $\sup_{\lambda\in [-\pi,\pi]}|\widehat{w}(\lambda)-w(\lambda)| \stackrel{P}{\rightarrow} 0$ and that
$\sup_{\lambda\in [-\pi,\pi]}|\widetilde{w}(\lambda)-w(\lambda)| \stackrel{P}{\rightarrow} 0$. To see (a) notice that, as in the proof of Proposition 4.3 $ (i) $,
\begin{align*}
\mbox{Var}^*(V^*_{1,n}) = & \frac{4\pi^2}{n}\sum_{j\in \mathcal G (n)}w(\lambda_{j,n}) \big(w(\lambda_{j,n})+w(-\lambda_{j,n})\big) f(\lambda_{j,n})^2\\
+ &  \frac{4\pi^2}{n}\sum_{j\in \mathcal G (n)}\Big(\widehat w(\lambda_{j,n})\big(\widehat w(\lambda_{j,n})+\widehat w(-\lambda_{j,n})\big) -w(\lambda_{j,n})\big(w(\lambda_{j,n})+w(-\lambda_{j,n})\big) \Big)\\
&\times \widehat{f}_n(\lambda_{j,n})^2 +\frac{4\pi^2}{n}\sum_{j\in \mathcal G (n)} w(\lambda_{j,n})\big(w(\lambda_{j,n})+w(-\lambda_{j,n})\big) \big(\widehat{f}_n(\lambda_{j,n})^2-f(\lambda_{j,n})^2\big).
\end{align*}
The first term on the right hand side of the above equality converges to $M(1,f)^4\cdot\tau^2_{1,R}$, while the second and the third converge to zero in probability, by Assumption \ref{assu_fnhat} and  $ \sup_{\lambda\in [-\pi,\pi]}|\widehat{w}(\lambda)-w(\lambda)| \stackrel{P}{\rightarrow} 0$.
To show (b) notice that by the definition of $\widehat{\sigma}_R^2$ and assertion (a), it suffices to show that $ \mbox{Var}^*(V_{2,n}^*) \stackrel{P}{\rightarrow} M(1,f)^4(\tau^2_{1,R} + \tau_{2,R})$ and that $ c_{n,R}\stackrel{P}{\rightarrow} M(1,f)^4\, \tau^2_{1,R}$. The former assertion was established in the proof of Theorem \ref{theorem1} $ (iii) $, while the latter follows using the same arguments as in the proof of Theorem \ref{theorem2} $ (i) $, and taking into account that $ \sup_{\lambda\in [-\pi,\pi]}|\widetilde{w}(\lambda)-w(\lambda)| \stackrel{P}{\rightarrow}  0$.  Finally, to establish (c) recall that $ V^*_{n,R}= V^*_{1,n}/D^*_n$ and that
$2\pi n^{-1}\sum_{j\in \mathcal G(n)}\widehat{f}_n(\lambda_{j,n}) \stackrel{P}{\rightarrow} M(1,f)$, while the bootstrap quantity $ \widetilde{D}_n^*=2\pi n^{-1}\sum_{j\in {\mathcal G}(n)}T^*(\lambda_{j,n})$ converges in probability to $ M(1,f) $ as well. The last convergence (for the doubly stochastic sequence $ \widetilde{D}_n^* $) means that
\bee
P^*(|\widetilde{D}_n^* - M(1,f)|> \varepsilon) =o_P(1) \quad \quad \forall \, \varepsilon >0\,,
\eee
and this assertion follows because  $ \mbox{E}^*(\widetilde{D}^*_n) \stackrel{P}{\rightarrow} M(1,f)$ and $ \mbox{Var}^*(\widetilde{D}_n^*)=\mathcal{O}_P(n^{-1})$, as can be seen from Proposition \ref{prop1} $ (i) $ when taking $ \varphi(\lambda)=1 $. Thus,
\bee
P^*(|D_n^* - M(1,f)^2|> \varepsilon) =o_P(1) \quad \quad \forall \, \varepsilon >0\,,
\eee
and it suffices to show that $ V^*_{1,n} $ is asymptotically normal with mean zero and variance $ M(1,f)^4 \, \tau^2_{1,R} $, that is,
\bee
\sup_{x \in \R} \Big| P^*( V_{1,n}^*\leq x ) - \Phi\Big( \frac{x}{M(1,f)^2 \, \tau_{1,R}} \Big) \Big|=o_P(1)\,.
\eee
For this we proceed as in the proof of Proposition \ref{prop1} $ (ii) $. Notice that $\mbox{E}^*(V_{1,n}^*)=0$ and write $ V_{1,n}^*=\sum_{j\in \mathcal G(n,+)} \big( \widetilde{Z}_{j,n}^*/\sqrt{n}\big)$ where
\[ \widetilde{Z}_{j,n}^* =2\pi(U^*_j-1)\big(\widehat w(\lambda_{j,n})\widehat{f}_n(\lambda_{j,n}) +\widehat w(-\lambda_{j,n})\widehat{f}_n(\lambda_{j,n}) \big), \quad \quad j\in \mathcal G (n,+)\,.\]
Lyapunov's condition is then verified in a straightforward way, along the lines of the proof of Proposition \ref{prop1} $ (ii) $ and using $\sup_{\lambda\in [-\pi,\pi]}|\widehat{w}(\lambda)-w(\lambda)| \stackrel{P}{\rightarrow} 0$. $ \hfill \square $


\end{document}